\documentclass[twocolumn, trackchanges]{aastex62}

\received{-}
\revised{-}
\accepted{-}

\shorttitle{IRX--$\beta$ relation and galaxy inclination at $z\sim1.5$}
\shortauthors{W. Wang et al.}

\begin{document}

\title{GALAXY INCLINATION AND THE IRX--$\beta$ RELATION: EFFECTS ON UV STAR-FORMATION RATE MEASUREMENTS AT INTERMEDIATE TO HIGH REDSHIFTS}

\correspondingauthor{Weichen Wang}
\email{wcwang@jhu.edu}

\author{Weichen Wang}
\affiliation{Department of Physics and Astronomy, Johns Hopkins University, 3400 N. Charles Street, Baltimore, MD 21218, USA}
\author{Susan A. Kassin}
\affiliation{Space Telescope Science Institute, 3700 San Martin Drive, Baltimore, MD 21218, USA}
\affiliation{Department of Physics and Astronomy, Johns Hopkins University, 3400 N. Charles Street, Baltimore, MD 21218, USA}
\author{Camilla Pacifici}
\affiliation{Space Telescope Science Institute, 3700 San Martin Drive, Baltimore, MD 21218, USA}
\author{Guillermo Barro}
\affiliation{Department of Physics, University of the Pacific, 3601 Pacific Avenue Stockton, CA 95211, USA}

\author{Alexander de la Vega}
\affiliation{Department of Physics and Astronomy, Johns Hopkins University, 3400 N. Charles Street, Baltimore, MD 21218, USA}

\author{Raymond C. Simons}
\affiliation{Department of Physics and Astronomy, Johns Hopkins University, 3400 N. Charles Street, Baltimore, MD 21218, USA}
\author{S. M. Faber}
\affiliation{UCO/Lick Observatory, Dept. of Astronomy and Astrophysics, University of California, Santa Cruz, CA 95064, USA}
\author{Brett Salmon}
\affiliation{Space Telescope Science Institute, 3700 San Martin Drive, Baltimore, MD 21218, USA}
\author{Henry C. Ferguson}
\affiliation{Space Telescope Science Institute, 3700 San Martin Drive, Baltimore, MD 21218, USA}
\author{Pablo G. P\'erez-Gonz\'alez}
\affiliation{Departamento de Astrof\'isica, Facultad de CC. F\'isicas, Universidad Complutense de Madrid, E-28040 Madrid, Spain}
\affiliation{Centro de Astrobiolog\'{\i}a (CAB, INTA-CSIC), Carretera de Ajalvir km 4, E-28850 Torrej\'on de Ardoz, Madrid, Spain}
\author{Gregory F. Snyder}
\affiliation{Space Telescope Science Institute, 3700 San Martin Drive, Baltimore, MD 21218, USA}
\author{Karl D. Gordon}
\affiliation{Space Telescope Science Institute, 3700 San Martin Drive, Baltimore, MD 21218, USA}

\author{Zhu Chen}
\affiliation{Shanghai Key Lab for Astrophysics, Shanghai Normal University, 100 Guilin Road, 200234, Shanghai, China}
\author{Dritan Kodra}
\affiliation{Department of Physics and Astronomy and PITT PACC, University of
	Pittsburgh, Pittsburgh, PA 15260, USA}

\begin{abstract}
At intermediate and high redshifts, measurements of galaxy star-formation rates are usually based on rest-frame ultraviolet (UV) data. A correction for dust attenuation, $A_\mathrm{UV}$, is needed for these measurements. This correction is typically inferred from UV spectral slopes ($\beta$) using an equation known as ``Meurer's Relation.'' In this paper, we study this relation at a redshift of 1.5 using images and photometric measurements in the rest-frame UV (\emph{HST}) through mid-infrared (\emph{Spitzer}). It is shown that massive star-forming galaxies (above $10^{10}\ M_\sun$) have dust corrections that are dependent on their inclination to the line-of-sight. Edge-on galaxies have higher $A_\mathrm{UV}$ and infrared excess (IRX=$L$(IR)/$L$(UV)) than face-on galaxies at a given $\beta$. Interestingly, dust corrections for low-mass star-forming galaxies do not depend on inclination. This is likely because more massive galaxies have more disk-like shapes/kinematics, while low-mass galaxies are more prolate and have more disturbed kinematics. To account for an inclination-dependent dust correction, a modified Meurer's Relation is derived: $A_\mathrm{UV}=4.43+1.99\beta - 1.73 (b/a-0.67)$, where b/a is the galaxy axis ratio. This inclination-dependence of $A_\mathrm{UV}$ can be explained by a two-component model of the dust distribution inside galaxies. In such a model, the dust attenuation of edge-on galaxies has a higher contribution from a “mixture” component (dust uniformly mixed with stars in the diffuse interstellar medium), and a lower contribution from a “birth cloud” component (near-spherical dust shells surrounding young stars in \ion{H}{2} regions) than that of face-on galaxies. The difference is caused by the larger path-lengths through disks at higher inclinations.
\end{abstract}

\keywords{dust, extinction -- galaxies: formation -- galaxies: high-redshift -- galaxies: star formation}

\section{Introduction} \label{sec:intro} 
For the study of galaxy formation at intermediate to high redshifts, it is common to measure star-formation rates (SFRs) from rest-frame ultraviolet (UV) luminosities, which are usually dominated by the light from young massive stars \citep{Conroy2013, Madau2014a}. UV-based SFRs need to be corrected for substantial dust attenuation (e.g., \citealt{Pannella2009a, Koprowski2017}). The dust correction for each galaxy is quantified by the UV dust attenuation, $A_\mathrm{UV}$, which is usually inferred from its UV spectral slope, $\beta$: $f(\lambda)\propto \lambda^{\beta}$, where $f$ is the flux density and $\lambda$ wavelength in the rest-frame UV \citep{Calzetti1994, Meurer1999}. The approach of measuring $A_\mathrm{UV}$ from $\beta$ is especially common at intermediate to high redshifts, where better or comparable star-formation rate indicators are seldom available (e.g., \citealt{Smit2012, Williams2014, Mehta2017, Santini2017}).

Therefore, if one wishes to measure integrated SFRs of galaxies from UV luminosities, they should constrain the $A_\mathrm{UV}$--$\beta$ relation \citep{Calzetti1997, Meurer1999}. One common way to do this in observations is to obtain far-infrared luminosities for a subsample of galaxies. The  ``infrared excess'' IRX, i.e., the infrared-to-UV luminosity ratio, is calculated as a robust proxy for $A_\mathrm{UV}$ \citep{Meurer1999, Gordon2000, Witt2000, Buat2012, Narayanan2018},  and the IRX--$\beta$ relation is measured \citep{Meurer1999}. The relation is then used to infer $A_\mathrm{UV}$ from $\beta$ and to calculate the SFRs of galaxies for which far-infrared data are unavailable.

However, the scatter of the IRX--$\beta$ relation is large (more than 0.5 dex, see e.g., \citealt{Grasha2013, Casey2014a, Reddy2017}). From a theoretical perspective, the scatter can be caused by variation in dust grain type, dust geometry, galaxy star-formation history, and galaxy stellar populations (e.g., \citealt{Charlot2000, Witt2000, Jonsson2010, Popping2017, Ferrara2017, Safarzadeh2017,Narayanan2018}). Observational studies have explored the IRX--$\beta$ relation as a function of galaxy age \citep{Kong2004, Reddy2010, Buat2012, Grasha2013}, SFR \citep{Nordon2013}, infrared luminosity \citep{Buat2005, Reddy2006a, Buat2012, Casey2014b}, stellar mass \citep{Reddy2017}, redshift \citep{Pannella2015, Reddy2017}, and dust optical depth (\citealt{Penner2012, Nordon2013, Forrest2016,Salmon2016}; Salim et al.\ 2018, submitted).

In spite of this progress, the effect of galaxy inclination to the line of sight on the IRX--$\beta$ relation  at intermediate to high redshifts has not been explored. Dust optical depth and the relative distribution of stars and dust along the line-of-sight vary with galaxy inclination, whereas the underlying stellar population does not. Therefore, inclination singles out the influence of the dust/star geometry and gives insight into details of the distribution of dust inside galaxies.

A few studies have shown indirect hints that the IRX--$\beta$ relation may be inclination-dependent at intermediate to high redshifts.  Observations of nearby galaxies find a correlation between galaxy inclination and the shape of the dust attenuation curve \citep{Wild2011,Chevallard2013,Battisti2017, Salim2018}. A theoretical study demonstrated that both $\beta$ and IRX vary strongly with the viewing angle \citep{Safarzadeh2017}. Recent work by \citet{Leslie2018} find that the $A_\mathrm{UV}$ value inferred from a relation $A_\mathrm{UV}=0.87(\beta+2.586)$ is overestimated for local face-on galaxies and underestimated for local edge-on galaxies, but such a trend is not found for $z\sim 0.7$ star-forming galaxies. However, their UV-selected sample at $z\sim 0.7$ may miss the most dusty edge-on galaxies due to the imaging sensitivity of \emph{Galaxy Evolution Explorer} (\emph{GALEX}).

With archival imaging by the \emph{Hubble Space Telescope} (\emph{HST}) Wide Field Camera 3 (WFC3) from the Cosmic Assembly Near-infrared Deep Extragalactic Legacy Survey (CANDELS, \citealt{Grogin2011, Koekemoer2011}), it is now feasible to measure galaxy inclination in the rest-frame optical for a large sample of galaxies. Along with deep 24\,\micron\ data from the Multiband Imaging Photometer for \emph{Spitzer} (MIPS) in the GOODS-South and GOODS-North fields (two of the five CANDELS fields), we are able to study the effect of galaxy inclination on the IRX--$\beta$ relation with a large sample of star-forming galaxies at intermediate redshifts. The redshift range we study in this work is $1.3<z<1.7$, which overlaps with the peak of cosmic star formation history ($1.5<z<2.0$, \citealt{Madau2014a}).

This paper aims at measuring and understanding the effect of galaxy inclination on the IRX--$\beta$ relation, the $A_\mathrm{UV}$--$\beta$ relation, galaxy dust attenuation curves, and UV-based SFR measurements at intermediate redshifts. It is organized as follows. \S \ref{sec:data} and \S \ref{sec:measurement} describe the observations and the measurement of physical quantities, respectively. \S \ref{sec:selection_cuts} describes the sample selection. The IRX--$\beta$ relation as a function of galaxy inclination for massive star-forming galaxies is presented in \S \ref{sec:rst_irxbeta}.  \S \ref{sec:meurerlawdiscuss} extends the discussion to low-mass galaxies and show that their  $A_\mathrm{UV}$--$\beta$ relation and IRX--$\beta$ relation may not vary with galaxy axis ratio. \S \ref{sec:rst_sfr} shows the effect of an inclination-dependent IRX--$\beta$ relation on UV-based SFR measurements, and \S \ref{sec:rst_irxbeta_interp} discusses the shape of UV dust attenuation curve as a function of galaxy inclination. \S \ref{sec:RT_mode} and \S \ref{sec:discussion_lowmass} use the results of radiative transfer models in the literature to understand the distribution of stars and dust inside star-forming galaxies. \S \ref{sec:conclude} presents the conclusions. Uncertainties in the measurements of $\beta$ and infrared luminosities, and the identification of compact galaxies are discussed in the Appendices.

A cold dark matter cosmology model with $\Omega_{\mathrm{M}}=0.3$ and $\Omega_{\mathrm{\Lambda}}=0.7$ is adopted, and the Hubble constant is set to 70 km s$^{-1}$\,Mpc$^{-1}$. All the magnitudes are in the AB magnitude system. We denote galaxy dust attenuation at rest-frame 1600\,\AA\ as $A_\mathrm{UV}$ throughout the paper.
\section{Data} \label{sec:data}
\begin{figure*}
	\centering
	\includegraphics[width=6.5in]{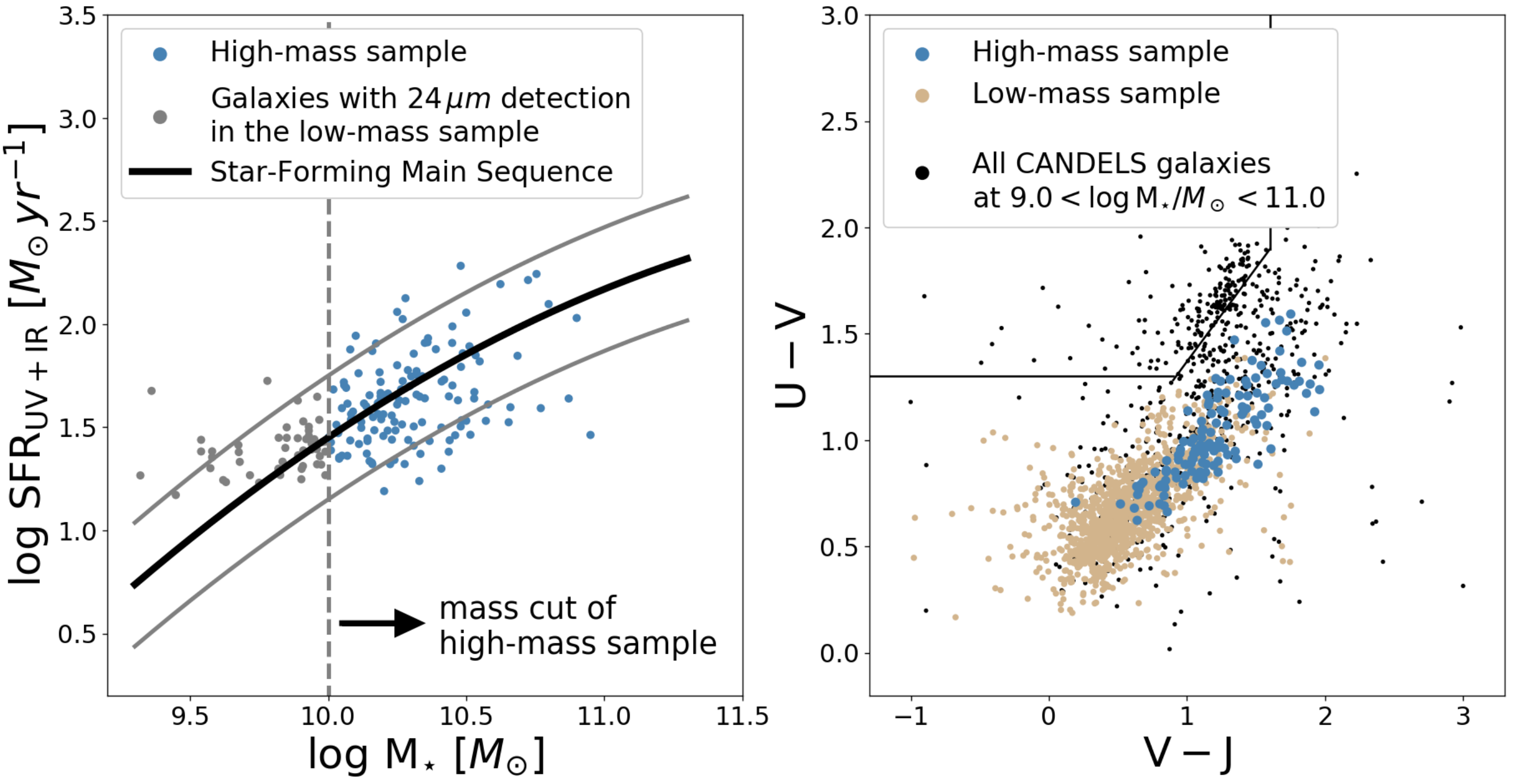}
	\caption{The galaxy sample is selected at $1.3<z<1.7$ and $9.0<\log\,M_\star/M_\sun<11.0$, and their general properties are presented in this plot. We create low-mass (brown points) and high-mass (black points) subsamples, whose stellar masses are below and above $10^{10}\,M_\sun$, respectively.	 The high-mass sample is representative of galaxies on the star-forming main sequence. Both samples have bluer ($\sim$0.5 mag) $\mathrm{U}-\mathrm{V}$ and $\mathrm{V}-\mathrm{J}$ colors than the general population of CANDELS galaxies at the same redshift and mass range (black points). 
	 \emph{Left}: On the SFR--$M_\star$ diagram, the high-mass sample (all with S/N$>$4 detection in 24\,\micron) stays close to the star-forming main sequence at $z=1.5$ \citep{Whitaker2014}, which is represented by the thick black line. The thinner gray lines mark 1-$\sigma$ (0.3 dex) deviations of the main sequence. Most galaxies in the low-mass sample do not have 24\,\micron\ fluxes to derive SFR$_\mathrm{UV+IR}$, so they are not shown. Those with 24\,\micron\ fluxes (gray points) in the low-mass sample stay significantly above the main sequence. \emph{Right}:  Solid lines separate quenched galaxies from the star-forming galaxies on the UVJ diagram, with the quenched galaxies lying in the upper left corner \citep{Williams2009}. Refer to \S  \ref{sec:selection_cuts} for more derails. \label{fig:sfms_uvj}}
\end{figure*}
We wish to study the effect of galaxy inclination on the IRX--$\beta$ relation and therefore on the measurement of UV SFR at intermediate redshifts. This requires a representative sample of star-forming galaxies with reliable redshifts and multi-band photometry from the rest-frame far-UV to at least mid-IR. The CANDELS GOODS-North and GOODS-South fields provide this multi-wavelength coverage over the redshift range $1.3 < z <  1.7$ (\citealt{Guo2013}; Barro et al.\ in prep.).

Two galaxy samples are used in this work. The high-mass sample is in the mass range $10.0<\log\,M_{\star}/M_{\sun}\,<11.0$. All of them have \emph{Spitzer} MIPS/24\,\micron\ fluxes. The low-mass sample is in the mass range $9.0<\log\,M_{\star}/M_{\sun}\,<10.0$. The majority of them do not have 24\,\micron\ fluxes to infer $L$(IR) and SFR$_\mathrm{UV+IR}$, due to the sensitivity of current mid-infrared dataset. The low-mass sample is discussed separately in this paper. 

Figure \ref{fig:sfms_uvj} shows the general properties of our selected samples with the SFR--$M_{\star}$ diagram and the rest-frame $\mathrm{U}-\mathrm{V}$ versus $\mathrm{V}-\mathrm{J}$ color diagram (UVJ diagram, \citealt{Williams2009}). The high-mass sample lies close to the star-forming main sequence (SFMS, e.g., \citealt{Brinchmann2004, Noeske2007, Whitaker2014}). Galaxies in the high-mass and low-mass sample are inside the star-forming region of the UVJ diagram, and have bluer ($\sim$0.5 mag) $\mathrm{U}-\mathrm{V}$ and $\mathrm{V}-\mathrm{J}$ colors than the general CANDELS galaxies in the same mass and redshift range.  

Details about the data, measurements, and sample selection follow in this section, \S\ref{sec:measurement}, and \S\ref{sec:selection_cuts} respectively. 
\subsection{UV, Optical, Near-IR, and Spitzer/IRAC}
Galaxies are selected from two of the five fields in the CANDELS survey, GOODS-South and GOODS-North, which have the deepest U-band data among the CANDELS fields. These U-band data probe the rest-frame 1600\,\AA\ to an absolute magnitude $M_{1600}$ of $-16.5$ mag at $z\sim1.5$. The U-band data of GOODS-South are from the Visible Multi-Object Spectrograph (VIMOS) on the Very Large Telescope (VLT, \citealt{Nonino2009, Guo2013}). The GOODS-North data are from the Large Binocular Cameras (LBC) on the Large Binocular Telescope (LBT, \citealt{Ashcraft2017}; Barro et al.\ in prep.).  The 5-$\sigma$ U-band limiting magnitudes of these data are 27.9 mag and 28.3 mag, respectively.

Details on the CANDELS photometry can be found in \citet{Guo2013}, and we give a brief review of important points below. In the CANDELS catalogs, photometry of \emph{HST}/ACS (F435W, F606W, F775W, F814W, F850LP), \emph{HST}/WFC3 (F105W, F125W, F160W), and \emph{Spitzer}/IRAC (Infrared Array Camera, \citealt{Fazio2004}) Channel 1--4 (3.6\,\micron, 4.5\,\micron, 5.8\,\micron, and 8.0\,\micron) is cross-matched (\citealt{Guo2013}, Barro et al.\ in prep.\,). The \emph{Spitzer}/IRAC data originally come from the GOODS \emph{Spitzer} Legacy project (PI: M. Dickinson), the \emph{Spitzer} Extended Deep Survey (SEDS; \citealt{Ashby2013}), and the \emph{Spitzer}-CANDELS survey (S-CANDELS; \citealt{Ashby2015}). Photometry of the \emph{HST} images is performed after all the bands are smoothed and matched to the WFC/F160W point spread function (PSF) using the {\sc{iraf/psfmatch}} algorithm. As for the ground-based U-band and \emph{Spitzer}/IRAC channels, the software {\sc{tfit}} \citep{Laidler2007} is used to perform photometry using the WFC3/F160W images as templates since they have high spatial resolution.
\subsection{24\,\micron}
\label{sec:data_mips} 
The MIPS/24\,\micron\ data are from the \emph{Spitzer} GTO project (Dickinson et al.\ 2003) and the GOODS survey (\citealt{Giavalisco2004}, also see \citealt{Perez-Gonzalez2008}). They reach a 5-$\sigma$ detection limit of 40\,$\mu$Jy (Barro et al.\ in prep.). Photometry was performed following \citet{Perez-Gonzalez2005} and \citet{Perez-Gonzalez2008} using the {\sc{ iraf/daophot}}  PSF-fitting algorithm. A radius of 7.5\arcsec\ was adopted to mitigate neighborhood contamination, and an aperture correction to the total flux is applied. The photometry is run in multiple passes as follows. The brightest sources are detected and subtracted first, and subsequently fainter sources afterwards. Such a strategy is suitable for intermediate-redshift extragalactic sources, given that the typical galaxy sizes (significantly smaller than $1\arcsec$ in radius) are considerably smaller than the PSF (3\arcsec\, in radius) and the fields are usually crowded. Each MIPS/24\,\micron\ source is then matched to the closest galaxy in the CANDELS catalog. 
To further eliminate confused sources, we refine the sample based on the proximity of neighbors as discussed in \S\ref{sec:selection_cuts}. 
\par 
Throughout this work, the 24\,\micron\ luminosity $L$(24\,\micron) is defined as $\nu L_\nu$ at the \emph{observed} 24\,\micron\ wavelength, where $\nu$ is the frequency and $L_\nu$ is the luminosity density per Hertz. It is close to the infrared luminosity at 8-10 \micron\ in the rest-frame.

\subsection{Far-Infrared}
Far-infrared imaging data (100\,\micron\ and 160\,\micron) are obtained with the \emph{Herschel} Photodetector Array Camera and Spectrometer (PACS, \citealt{Poglitsch2010}).  Observations were conducted by the following surveys in the GOODS fields: the PACS Evolutionary Probe (PEP; \citealt{Berta2011,Lutz2011}), GOODS-\emph{Herschel} (\citealt{Elbaz2011,Magnelli2013}), and the \emph{Herschel} Multi-tiered Extragalactic Survey (HerMes; \citealt{Oliver2012}). Details about the photometry and data reduction process are described in \citet{Perez-Gonzalez2010} and Barro et al.\ (in preparation). Our far-infrared photometry uses \emph{Spitzer} MIPS/24\,\micron\ images as position priors and the {\sc{iraf/daophot}} PSF-fitting algorithm. To avoid source confusion, \emph{Herschel} detections are matched to the MIPS/24\,\micron\ sources only if there is a unique MIPS object within the PACS PSFs (3.4\arcsec\ and 5.5\arcsec\ in radius for PACS/100\,\micron\ and PACS/160\,\micron, respectively). 

\subsection{X-ray}
\label{sec:Xray}
The Chandra X-ray catalogs in GOODS-South (4 million-second depth) and GOODS-North (2 million-second depth) from \citet{Xue2011,Xue2016} are adopted to identify X-ray AGNs, which are later removed in the sample selection process. The X-ray sources are matched with the CANDELS catalogs with a radius of 1.5\arcsec. The flux limits of these datasets are $3.2\times10^{-17}$ erg$\mathrm{s}^{-1}$cm$^{-2}$ and 3.5$\times 10^{-17}$ erg$\mathrm{s}^{-1}$cm$^{-2}$ at 0.5--7 keV for GOODS-South and GOODS-North, respectively.

\section{Measurements}
\label{sec:measurement}
\subsection{Redshifts, Stellar Masses, and Central Stellar Mass Surface Densities}
\label{sec:measurement_redshift}
GOODS-South and GOODS-North are extensively covered by major spectroscopic surveys at around $z=1.5$, such as DEEP2 \citep{Newman2013}, VVDS/VUDS \citep{LeFevre2013, LeFevre2015}, MOSDEF \citep{Kriek2015}, 3D-HST \citep{Morris2015, Momcheva2016}, as well as many others (e.g.,\ \citealt{Reddy2006b, Barger2008,Hathi2009, Cooper2012,Kurk2013,Pirzkal2013,Wirth2015}). We adopt a compilation of the existing redshift catalogs by the CANDELS team (Kodra et al.\ in prep.). A total of 81 galaxies have spectroscopic redshifts at $10.0<\log\,M_\star/M_\sun<11.0$ (63.0\% of the final sample; 54 from 3D-HST, 14 from MOSDEF, and 6 from VLT-based surveys). 

Photometric redshifts are used for the rest of the galaxies. They are obtained based on the fit to the CANDELS multi-wavelength catalogs from U-band to IRAC Channel 4 (8.0\,\micron) by Kodra et al.\ (in prep.). For each galaxy,  five independent codes are used to derive photometric redshifts \citep{Bolzonella2000,Fontana2000,Brammer2008,Wiklind2008,2011ascl.soft08009A}. The median of the resulting photo-z values is derived from the output probability distribution functions.
A comparison with spectroscopic redshifts shows that these CANDELS photometric redshifts have an average accuracy of 0.02 (Kodra et al.\ in prep.). Using these spectroscopic and photometric redshifts, the CANDELS rest-frame color catalog (P.I.: D. Kocevski, S. Wuyts and G. Barro) was derived using the {\sc{eazy}} code \citep{Brammer2008} and is adopted in this work. 

We use the so-called CANDELS ``reference" mass in this work, which is calculated as follows.  For each galaxy in the CANDELS catalog, the stellar masses are derived independently from spectral energy distribution (SED) fitting by 10 teams, who used the same set of photometric data from the U-band to the IRAC channels (\citealt{Guo2013}; Barro et al.\ in prep.) but adopted different assumptions regarding the star-formation history, dust attenuation law, and stellar metallicity of galaxies and the fitting method \citep{Santini2015}. The ``reference mass'' is the median of these mass values. Stellar masses calculated in this manner reduce the systematic biases caused by any specific assumption in the SED-fitting. 
\par 
Measurements of the stellar mass surface density within the central 1 kpc of galaxies (0.37\arcsec\ at $z=1.5$), $\Sigma_{1\mathrm{kpc}}$, are adopted from Chen et al.\ (in prep.). They were measured by running the SED-fitting code {\sc{fast}} \citep{Kriek2009} on the 9 \emph{HST} bands in the CANDELS multi-wavelength aperture photometry catalog (Liu et al.\ in prep.\,), following the same procedures and assumptions as \citet{Barro2017}. These $\Sigma_{1\mathrm{kpc}}$ values are used to identify massive compact galaxies in Appendix \ref{sec:appendix_morp}. 
\subsection{Galaxy Inclinations and Identification of Compact Galaxies}
\label{sec:measurement_incli}
Galaxies are studied in two separate stellar mass ranges in this work, $10.0<\log\,M_\star/M_\sun<11.0$ and $9.0<\log\,M_\star/M_\sun<10.0$, respectively.

For the star-forming galaxies in the range $10.0<\log\,M_\star/M_\sun<11.0$, which we refer to as the high-mass sample, we classify their morphology into two types: disk galaxies and compact galaxies. This classification is based on their S\'ersic indexes \citep{Sersic1963} measured from their \emph{HST} WFC3/F160W images. Simply, galaxies with S\'ersic indexes larger than 2.0 are classified as compact galaxies, and the rest are disk galaxies. This classification is examined in Appendix \ref{sec:appendix_morp}. The compact galaxies defined in this way have high  stellar mass density in the central 1 kpc region, $\Sigma_{1\mathrm{kpc}}$, which is consistent with the definition of compact galaxies according to their $\Sigma_{1\mathrm{kpc}}$ by \cite{Barro2017}.  

For disk galaxies, the ratio between the minor and major axes of galaxies ($b/a$) is used as a measure of their inclination to the line of sight. The axis ratios used in this work are measured with the {\sc{galfit}} morphological code \citep{Peng2002} on the CANDELS WFC3/F160W images \citep{VanderWel2012}. We place disks into two broad categories: those with $b/a>0.5$ which we refer to as ``face-on'', and those with $b/a<0.5$ which we refer to as ``edge-on.''  The inclination angle $i$ of disk galaxies can be calculated following \cite{Holmberg1958}:
\begin{equation}
\cos^2 i = \frac{(b/a)^2-\alpha^2}{1-\alpha^2},
\end{equation}
 where $\alpha$ is the ratio between disk scale height and scale length. The value of $\alpha$ is set to be 0.2 following \cite{Simons2016}, which is within the range found for local galaxies, $0.1<\alpha<0.3$ \citep{Padilla2008,Unterborn2008,Rodriguez2013}.  For  $\alpha=0.2$, the two ranges of axis ratio, $b/a<0.5$ and $b/a>0.5$, correspond to $0^\circ<i<62^\circ$ and  $62^\circ<i<90^\circ$, respectively. There are 69 and 49 galaxies in each axis ratio bin, respectively, for our high-mass sample. Since we adopt broad ranges of axis ratio, categorization of galaxies based on their inclination angles is not sensitive to the adopted $\alpha$ value in the above conversion relation.

As for the low-mass galaxy sample ($9.0<\log\,M_\star/M_\sun<10.0$), their shapes can be either prolate or oblate \citep{VanderWel2014, Zhang2018} and do not have disk-like kinematics \citep{Kassin2012,Simons2017}. Hence, we do not use the term ``inclination'' to describe them, and simply classify them as high and low axis-ratio galaxies ($b/a>0.5$ and $b/a<0.5$, respectively, refer to Appendix \ref{sec:appendix_morp}) throughout the paper. We discuss their intrinsic shapes further in \S \ref{sec:discussion_lowmass}.
\subsection{UV Spectral Slopes ($\beta$)}

\label{sec:measurement_beta}
The UV spectral slope of a galaxy, or $\beta$, is defined as $f(\lambda)\propto \lambda^{\beta}$, where $f$ is the flux density and $\lambda$ is the wavelength in the rest-frame UV \citep{Calzetti1994, Meurer1999}. We use three wavebands to measure $\beta$: VIMOS and LBC U-band, ACS/F435W, and ACS/F775W. This corresponds to 1450-3070\,\AA\ in the rest-frame at $z=1.5$. Note that ACS/F606W is not used, which corresponds to 1940-2850\,\AA\ in the rest-frame at $z=1.5$, because it falls too close to the rest-frame 2175\,\AA\ bump feature in the dust attenuation curves. Including this band may bias the measurement of the UV spectral slope (e.g., \citealt{Popping2017,Tress2018}).
\par 
To obtain $\beta$, a power-law is fit to the fluxes in these three bands (U-band, ACS/F435W, and ACS/F775W) with the least-square method. The reciprocals of the uncertainties on the photometry serve as weights in the fitting. Before the fitting, a Galactic extinction correction for each filter is applied \citep{Schlafly2011}. These corrections influence the final values of $\beta$ by $\sim\,$0.025 for GOODS-South and $\sim\,$0.037 for GOODS-North. 
\par 
Model tests are run to investigate potential systematic uncertainties in the measurement of $\beta$ in Appendix \ref{sec:appendeix}. The resulting $\beta$ values are on average redder than the spectroscopically measured $\beta$ values \citep{Calzetti1994} by 0.1-0.3, which is shown in Figure \ref{fig:beta_sedfittingtest}. Such systematic error is not correlated with galaxy inclination if face-on and edge-on galaxies do not have the 2175\,\AA\ bump in their dust attenuation curves. The influence of a 2175\,\AA\ bump, which may vary with galaxy inclination, is smaller than 0.1 in $\beta$ according to the tests with model galaxy spectra in Figure \ref{fig:deltabeta_beta}. 

\subsection{Total Infrared Luminosity $L$(IR)}

\begin{figure}
	\centering
	\includegraphics[width=3.2in]{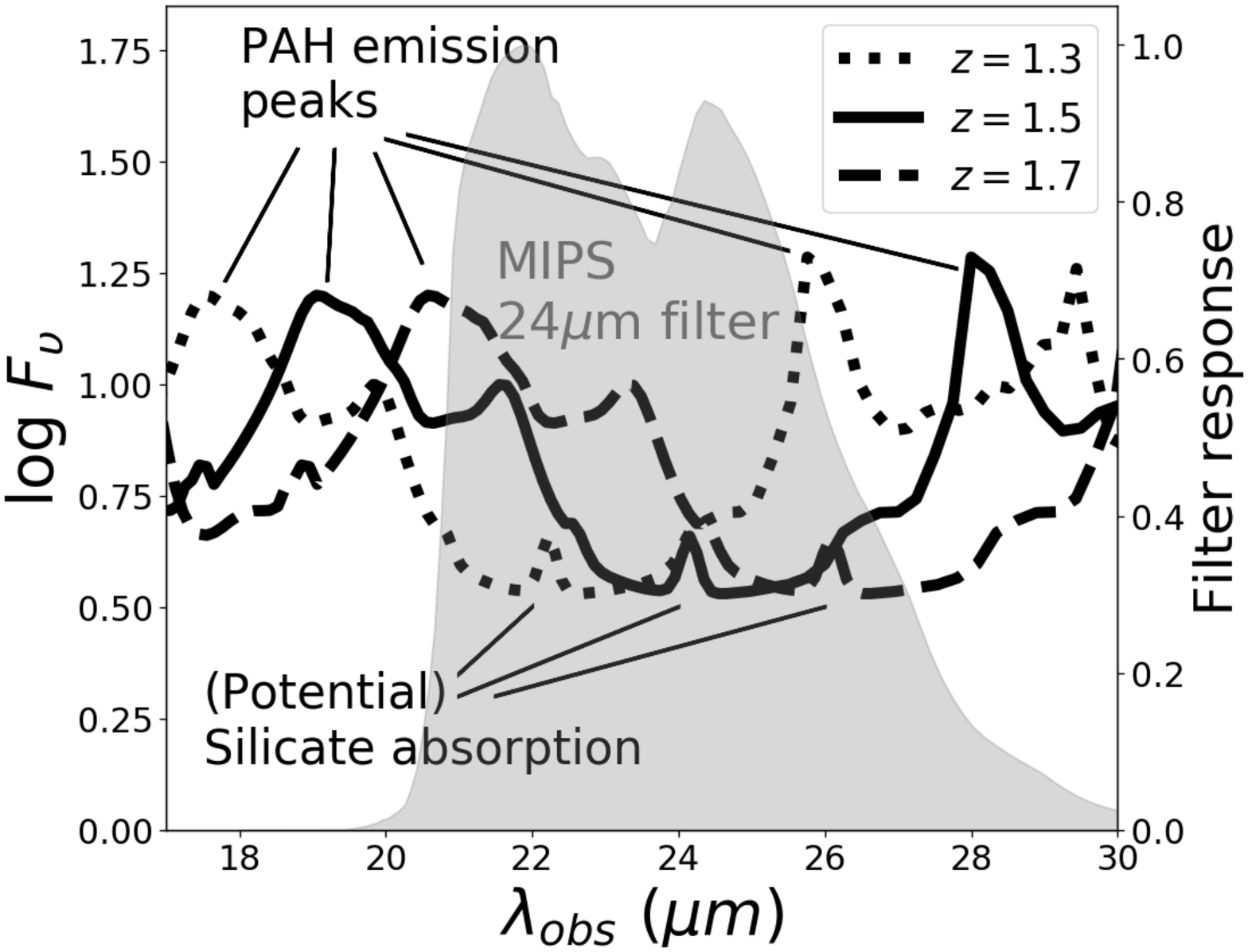}
	\caption{Infrared spectral templates for local galaxy with  $L$(IR)=$10^{10.75}L_{\sun}$ \citep{Rieke09}  are shown as black curves. They are plotted in the observed frame and have arbitrary normalizations. In the redshift range of this study ($1.3<z<1.7$), the MIPS/24\,\micron\ passband is used to infer the total infrared luminosity $L$(IR). Its response curve is shown as the gray shadow in the plot.  It falls between two major PAH emission peaks (rest-frame 8.6\,\micron\ and 11.3\,\micron ) and overlaps with the rest-frame 9.7\,\micron\ silicate absorption feature.  This means that the observed 24\,\micron\ luminosity is dominated by hot dust emission, PAH features, and potential silicate absorption. \label{fig:spec_filters}}
\end{figure}
In this paper, $L$(IR) is defined as the integrated luminosity from rest-frame 8\,\micron\, to 1000\,\micron. Ideally, far-infrared photometry would be used to derive $L$(IR), given that the peak of cold dust emission is at approximately 100\,\micron\ in the rest-frame. The \emph{Herschel} telescope is the only realistic public source of such data. However, due to its sensitivity and spatial resolution (6.7\arcsec\ at 100\,\micron, 11\arcsec\ at 160\,\micron\, and 17.6\arcsec\ at 250\,\micron), stacking is generally required for galaxies like those in our sample, and image confusion can be a major source of error.  Therefore, in the main part of this paper, we choose to infer $L$(IR) from the \emph{Spitzer}/24\,\micron\ flux. \emph{Spitzer}/MIPS has a sensitivity which is at least 3 times greater than \emph{Herschel} (see Fig.\ 1 of \citealt{Madau2014a}), and its PSF is slightly smaller (6\arcsec), making individual detection possible. 

To infer the $L$(IR) of galaxies from the 24\,\micron\ fluxes, we adopt the conversion from \citet{Rujopakarn2013}. The method is based on two arguments. First, it is assumed that the star-forming area and IR emission regions are generally extended across the entire galaxy at intermediate to high redshifts. Second, it is assumed that galaxy SEDs from 5\,\micron\ to 1000\,\micron\ are well characterized by the surface density of $L$(IR) at 1$<z<$3 \citep{Rujopakarn2011}.  The $L$(IR) values given by \cite{Rujopakarn2013} are found to be consistent with those obtained by fitting the \emph{Herschel} far-IR photometry \citep{Salmon2016}. However, compared to another widely-used $L$(IR)--$L$(24\,\micron) relation by \cite{Wuyts2008}, the $L$(IR) given by the \cite{Rujopakarn2013} relation is around 0.15 dex fainter \citep{Salmon2016}. If the conversion of \cite{Wuyts2008} is adopted, the conclusions of this paper do not qualitatively change. 

Besides this potential systematic discrepancy, there is a drawback of using MIPS/24\,\micron\ to infer $L$(IR) at $z\sim 1.5$, namely the silicate absorption. In Figure \ref{fig:spec_filters}, we show how a galaxy SED is sampled by the 24\,\micron\ passband at $1.3<z<1.7$. The MIPS response curve, shown as the gray shaded area, lies between the 8.6\,\micron\ and 11.3\,\micron\ Polycyclic Aromatic Hydrocarbon (PAH) emission peaks in the galaxy SEDs, so $L$(24\,\micron) is dominated by hot dust emission, PAH features, and unfortunately also the 9.7\,\micron\ silicate absorption feature. 
The silicate absorption can bring potential systematic uncertainties to the derived $L$(IR) because it attenuates the observed 24\,\micron\ flux and varies with silicate optical depth along the line of sight and therefore with galaxy inclination (see \citealt{Jonsson2010}). We argue in Appendix \ref{sec:24umuncertainty} that the uncertainty in the $L$(IR) measurement due to silicate absorption is 0.08 dex at most, and do not qualitative influence the main conclusions presented in this paper. 

Appendix \ref{sec:24umuncertainty} presents more details about $L$(IR) uncertainties and compares the $L$(IR) values converted from 24\,\micron\ with those converted from additional \emph{Herschel} far-IR fluxes. The difference between the two kinds of $L$(IR) values are typically smaller than 0.2 dex. This is in agreement with other studies \citep{Elbaz2011, Rujopakarn2013,Salmon2016}. 
\begin{figure*}
	\centering
	\includegraphics[width=6.5in]{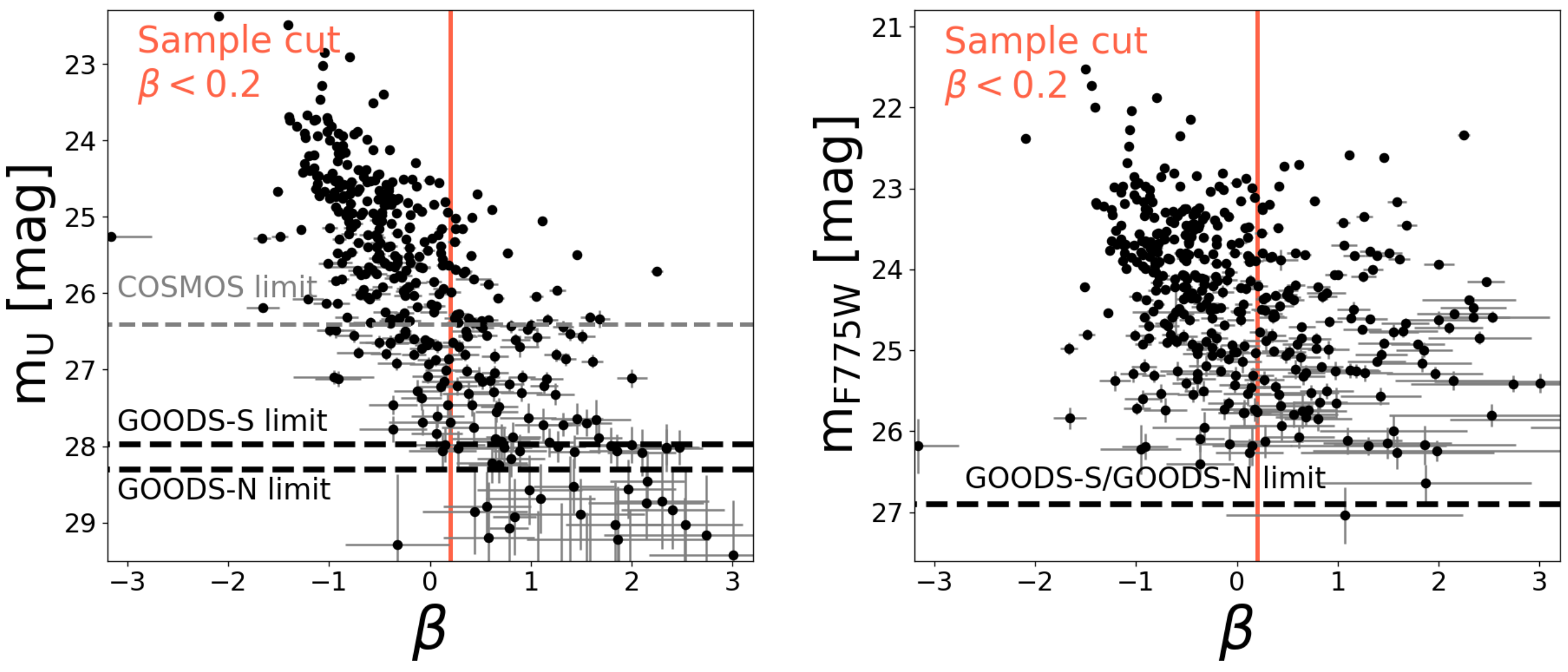}
	\caption{To keep sample galaxies above the detection limits in both the ground-based U-band (left) and \emph{HST} ACS/F775W (right), a selection cut is applied to the UV spectral slope $\beta$ . This figure shows the magnitudes versus $\beta$ plots for the high-mass sample ($10.0<\log\,M_\star/M_\sun<11.0$).  \textit{Left}: The observed U-band magnitude is correlated with $\beta$. The two black dashed lines mark the 5-$\sigma$ detection limits of the U-band data in GOODS-South and GOODS-North. As $\beta$ increases, galaxies with $\beta>0.2$ start to fall below the detection limits. For reference, the U-band limit of previous work using COSMOS-CFHT data \citep{Heinis2013} is shown as a gray dashed line. \textit{Right}: No strong correlation is found between $\mathrm{m}_{\mathrm{F775W}}$ and $\beta$, and most galaxies lie above the 5-$\sigma$ detection limits in ACS/F775W for GOODS-South and GOODS-North \citep{Skelton2014}. \label{fig:UVslopes}}
\end{figure*} 

\subsection{Star-Formation Rates and UV Luminosities}
\label{sec:measurement_sfr}
Two SFR indicators are used in this paper. Both assume a \citet{Chabrier2003} initial mass function. The first SFR indicator uses rest-frame UV and IR photometry following \citet{Whitaker2014}:
\begin{equation}
\mathrm{SFR}_{\mathrm{UV+IR}}=1.09\times 10^{-10}[L(\mathrm{IR})+3.3L(2800\,\mbox{\AA}\,)], \label{eqn:sfr_uvir}
\end{equation}
where  the unit of $\mathrm{SFR}_{\mathrm{UV+IR}}$ is $M_\sun\,yr^{-1}$, and the unit of the luminosities, $L$(IR) and $L(2800\,\mbox{\AA}\,)$, is $L_{\sun}$. The second SFR indicator is purely based on the rest-frame UV corrected for dust following \citet{Kennicutt2012}:
\begin{equation}
\mathrm{SFR}_{\mathrm{UV,\,\beta}}=1.72\times 10^{-10}L(1600\,\mbox{\AA}\,)\cdot 10^{0.4\,A_{\mathrm{UV,\,Meurer}}}, \label{eqn:sfrbeta}
\end{equation}
where the units of SFR and luminosity are the same as in Equation 1. The quantity $A_\mathrm{UV,\,Meurer}$ is the dust attenuation in magnitude at rest-frame 1600\,\AA\ and is derived from the UV spectral slope $\beta$ following \citet{Meurer1999}:
\begin{equation}
A_\mathrm{UV,\,Meurer}=4.43+1.99\,\beta. \label{eqn:meurerrelation}
\end{equation}
The luminosities $L$(1600\,\AA) and $L$(2800\,\AA) in the equations above are UV luminosities defined as $\nu L_\nu$ at rest-frame 1600\,\AA\ and rest-frame 2800\,\AA, respectively, where $\nu$ is the frequency and $L_\nu$ is the luminosity density per Hz. 

\section{Sample Selection}
\label{sec:selection_cuts}
Two galaxy samples are used in this work:
\begin{enumerate}
	\item A high-mass sample spanning stellar masses $10.0<\log\,M_\star/M_\sun<11.0$. All of these galaxies have reliable 24\,\micron\ fluxes.
	\item A low-mass sample  spanning $9.0<\log\,M_\star/M_\sun<10.0$. Most of these galaxies do not have 24\,\micron\ fluxes.
\end{enumerate}
Both samples are selected from the GOODS-South and GOODS-North fields. The mass range of high-mass galaxies is determined to ensure that galaxies with 24\,\micron\, fluxes approximately trace the SFMS and do not lie too far above it (see Figure\ \ref{fig:sfms_uvj}). No selection related to the 24\,\micron\ fluxes is applied to the low-mass galaxies, considering that most star-forming galaxies in this mass range fall below the \emph{Spitzer} MIPS/24\,\micron\ detection limit. 

A full list of the selection cuts applied to the CANDELS catalog is (\citealt{Guo2013}; Barro et al.\ in prep.) is as follows:
\begin{enumerate}
	\item Galaxies are selected in the redshift range $1.3<z<1.7$, where the U-band data probe the rest-frame far-UV and spectroscopic redshifts are abundant (\S\ref{sec:measurement_redshift}). Stars, artifacts, and catastrophic errors in the photometry are identified by requiring the Sextractor \citep{Bertin1996} parameter CLASS\_STAR to be smaller than 0.9 and the CANDELS photometry flags to be 0. This cut leaves a total of 520 galaxies in the high-mass sample and 1651 galaxies in the low-mass sample.
	\item AGNs are removed using the X-ray criteria \citep{Xue2011, Xue2016} and the \emph{Spitzer}/IRAC color criteria \citep{Donley2012}. The X-ray AGN criteria are based on the observed X-ray luminosity, effective photon index, X-ray-to-optical flux ratio, and spectroscopic features, which are detailed in \S 4.4 of \cite{Xue2011}. The IRAC color criteria of \citet{Donley2012} are as follows:
	\begin{eqnarray}
	 && f_\mathrm{8.0\,\micron}>f_\mathrm{5.8\,\micron}>f_\mathrm{4.5\,\micron}>f_\mathrm{3.6\,\micron},\nonumber\\
	&& \log\,(f_\mathrm{8.0\,\micron}/f_\mathrm{4.5\,\micron})- 1.21\times\log\,(f_\mathrm{5.8\,\micron}/f_\mathrm{3.6\,\micron})\nonumber\\
	&&\qquad\qquad\qquad\qquad\qquad =[-0.27, +0.27] ,\nonumber\\
	&& \log\,(f_\mathrm{8.0\,\micron}/f_\mathrm{4.5\,\micron})\geqslant 0.15,\nonumber\\
	&& \log\,(f_\mathrm{5.8\,\micron}/f_\mathrm{3.6\,\micron})\geqslant 0.08,
	\end{eqnarray}
	where $f_\mathrm{3.6\,\micron}$, $ f_\mathrm{4.5\,\micron}$, $f_\mathrm{5.8\,\micron}$, and $f_\mathrm{8.0\,\micron}$ are the observed fluxes in IRAC Channel 1-4, respectively. 
	The cuts remove 56 and 13 galaxies from the high-mass sample and the low-mass sample, respectively.  We are aware of the drawback that probably not all AGNs are removed, due to the sensitivities of available X-ray and mid-IR dataset and the obscuration of AGN by gas and dust. Such drawback is unavoidable with current facilities (see a review by \citealt{Hickox2018}). 
	\item A cut in $\beta$ is applied to preserve sample completeness, because the observed U-band magnitude (rest-frame far-UV) is correlated with $\beta$ and falls below the detection limit at high $\beta$ values (Figure \ref{fig:UVslopes}). We require $\beta<0.2$ for the high-mass sample, and for the low-mass sample we apply $\beta<-0.5$ and $\beta<0.0$ to stellar mass ranges $9.0<\log\,M_\star/M_\sun<9.5$ and $9.5<\log\,M_\star/M_\sun<10.0$, respectively. The cut reduces the high-mass sample to 219 galaxies and the low-mass sample to 1332 galaxies. 
	
	As is shown in Figure \ref{fig:UVslopes}, the UV slope $\beta$ is only weakly correlated with the ACS/F775W magnitude, but is strongly correlated with the U-band magnitude. Galaxies start to fall below the 5-$\sigma$ U-band detection limit at $\beta=0.2$, while always stay above the detection limit on the right panel. Beyond the UV slope range we select, U-band faint galaxies, which have probably high dust attenuation, can be missing from the sample. Trends with the low-mass sample are similar to the high-mass sample and the cuts in $\beta$ are set in the same way as described here.

	\item For the high-mass sample only, an additional selection is made to ensure reliable 24\,\micron\ photometry. First, the 24\,\micron\ flux is required to be higher than 30.0 $\mu$Jy  to ensure a high enough signal-to-noise ratio  (S/N$>4.0$) and to remove objects with strong silicate absorption at rest-frame 9.7\,\micron, which is typical for deeply obscured AGNs (e.g., \citealt{Levenson07, Hickox2018}). Then, only the galaxies that are well-isolated in the \emph{Spitzer} MIPS/24\,\micron\, images are selected to avoid serious confusion with neighboring sources as per two criteria. The first is that there should be no more than one F160W source inside a circular aperture with the size of the MIPS/24\,\micron\ PSF (3\arcsec\ in radius). The second is that the separation between the MIPS source center and the F160W counterpart must be less than 1\arcsec. The flux cut removes 45 galaxies and the following two cuts to avoid source confusion remove 41 galaxies, leaving 133 in the high-mass sample. We discuss whether this selection can bias the main conclusions of this paper in \S \ref{sec:rst_irxbeta_rst}.
	\item Quenched galaxies are removed based on galaxy positions on the rest-frame $\mathrm{U-V}$ vs.\ $\mathrm{V-J}$ diagram (UVJ diagram, \citealt{Williams2009}). This removes only 1 galaxies from the high-mass sample and 11 galaxies from the low-mass sample. 
	\item We inspect the WFC3/F160W images of the remaining galaxies and remove visually identified major mergers. Objects with one or more major contacting companions within a distance of 2\arcsec\ (16.7 kpc at $z=1.5$) are identified as mergers. This selection removes 3 galaxies in the high-mass sample and 56 in the low-mass sample.
 \end{enumerate}

\begin{figure*}
	\centering
	\includegraphics[width=4.5in]{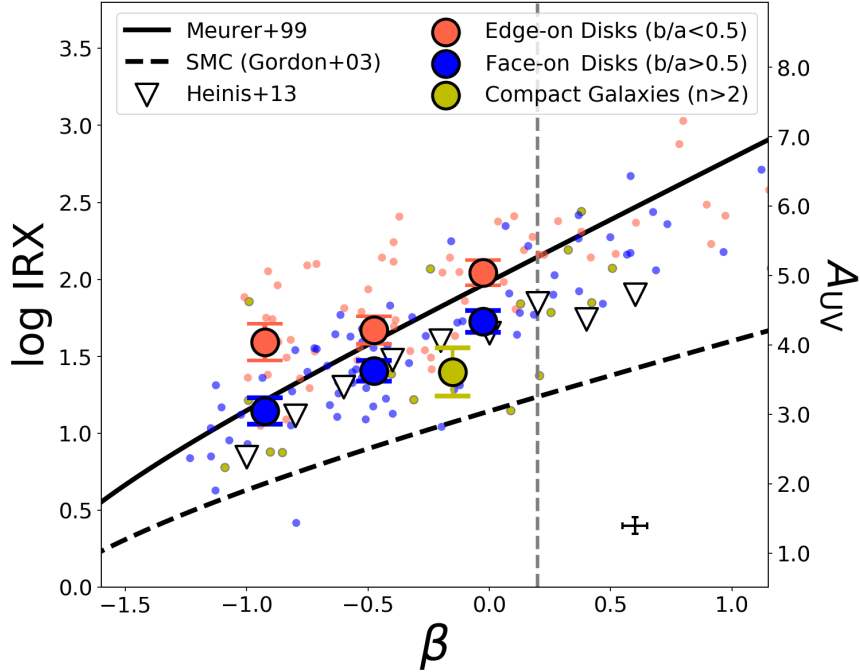}
	\caption{For the high-mass sample ($10.0<\log\,M_\star/M_\sun<11.0$), edge-on galaxies have greater values of IRX (=$L$(IR)/$L(1600\,\mbox{\AA})$, left axis), and effectively higher $A_\mathrm{UV}$ (right axis), than face-on galaxies at the same UV spectral slope $\beta$. Small points show individual galaxies and large circles indicate the medians in equal-spaced bins of $\beta$ from $\beta=-1.15$ to $\beta=0.20$. Errorbars on the median points indicate their standard deviations. The errorbar on the bottom right indicates the typical uncertainty of individual galaxies. A vertical dashed gray line indicates the $\beta$ value beyond which the sample is incomplete (Figure \ref{fig:UVslopes}). The local starburst relation \citep{Meurer1999}, the dust law for the Small Magellanic Cloud (SMC, \citealt{Gordon2003}), and a previous measurement for $z\sim1.5$ galaxies \citep{Heinis2013} are also plotted.  In this figure, $A_\mathrm{UV}$ on the right vertical axis is calculated  from IRX following equation (1) of \citet{Overzier2011}, rather than from $\beta$. \label{fig:irx_beta}}
\end{figure*}

The final sample used in this paper consists of 129 galaxies in the high-mass sample and 1265 in the low-mass sample. The general properties of these samples are shown in Figure \ref{fig:sfms_uvj}. On the SFR--$M_\star$ diagram (left panel), the high-mass sample is representative of the SFMS at $z=1.5$. The low-mass sample is not shown, because most of them do not have 24\,\micron\ fluxes to infer SFR$_\mathrm{UV+IR}$, due to the sensitivity of \emph{Spitzer} MIPS/24\,\micron. On the UVJ diagram (right panel), all the sample galaxies are in the star-forming region. Both the high-mass sample and the low-mass sample have bluer ($\sim$0.5 mag) $\mathrm{U-V}$ and $\mathrm{V-J}$ colors compared to the general star-forming galaxies in the same mass and redshift range. This is due to the $\beta$ cut (Criterion 4 above).

\section{Effect of Galaxy inclination on the IRX--$\beta$ relation and the $A_\mathrm{UV}$--$\beta$ relation}
\label{sec:rst_irxbeta}

This section examines the effect of galaxy inclination (or axis ratio) on the IRX--$\beta$ relation and the $A_\mathrm{UV}$--$\beta$ relation. Dust attenuation values ($A_\mathrm{UV}$) of galaxies with different inclinations are compared at a given UV spectral slope $\beta$. 
We show that in general edge-on galaxies have higher $A_\mathrm{UV}$ and IRX than face-on galaxies at a given $\beta$ for the high-mass sample, whereas the $A_\mathrm{UV}$--$\beta$ relation likely does not vary with galaxy axis ratio for the low-mass sample.

The IRX--$\beta$ relation for the high-mass sample is measured in \S\ref{sec:rst_irxbeta_rst}. This diagram links the relative amount of UV emission and IR dust emission (i.e., IRX=$L$(IR)/$L$(1600\,\AA)\,) with the UV slope $\beta$.  Dust attenuation $A_\mathrm{UV}$ can be directly converted from IRX. For star-forming galaxies without 24\,\micron\ data, their $L$(IR) and IRX can not be derived. A separate discussion which makes use of the UV luminosity versus $\beta$ relation is presented in \S \ref{sec:rst_irxbeta_rst_wo24}.

For the the low-mass sample, most of the galaxies do not have 24\,\micron\ detection. We take the same approach as \S \ref{sec:rst_irxbeta_rst_wo24} and use the UV luminosity versus $\beta$ relation to show that galaxies with different axis ratios in the sample may have the same $A_\mathrm{UV}$  at a given $\beta$ (\S\ref{sec:meurerlawdiscuss}).

The \citet{Meurer1999} IRX--$\beta$ relation adopted in this paper is slightly different from its original form due to a different definition of IRX.  IRX was defined as the ratio between far-IR luminosity (rest-frame 42.5-122.5\,\micron) and far-UV luminosity (rest-frame 1600\,\AA) in \citet{Meurer1999}, whereas in this work we use the total infrared luminosity (rest-frame 8--1000\,\micron) instead of the far-IR luminosity. This total IR luminosity is inferred from the observed 24\,\micron\ flux following \cite{Rujopakarn2013}. To account for the difference, we replace the bolometric correction factor BC(FIR)=1.4 with BC(IR)=1.0 in the original derivation of IRX  by \citet{Meurer1999} following \cite{Overzier2011}. This increases IRX by 0.15 dex with respect to the original equation (eqn.\ 10 of \citealt{Meurer1999}).

\begin{figure*}
	\centering
	\includegraphics[width=7in]{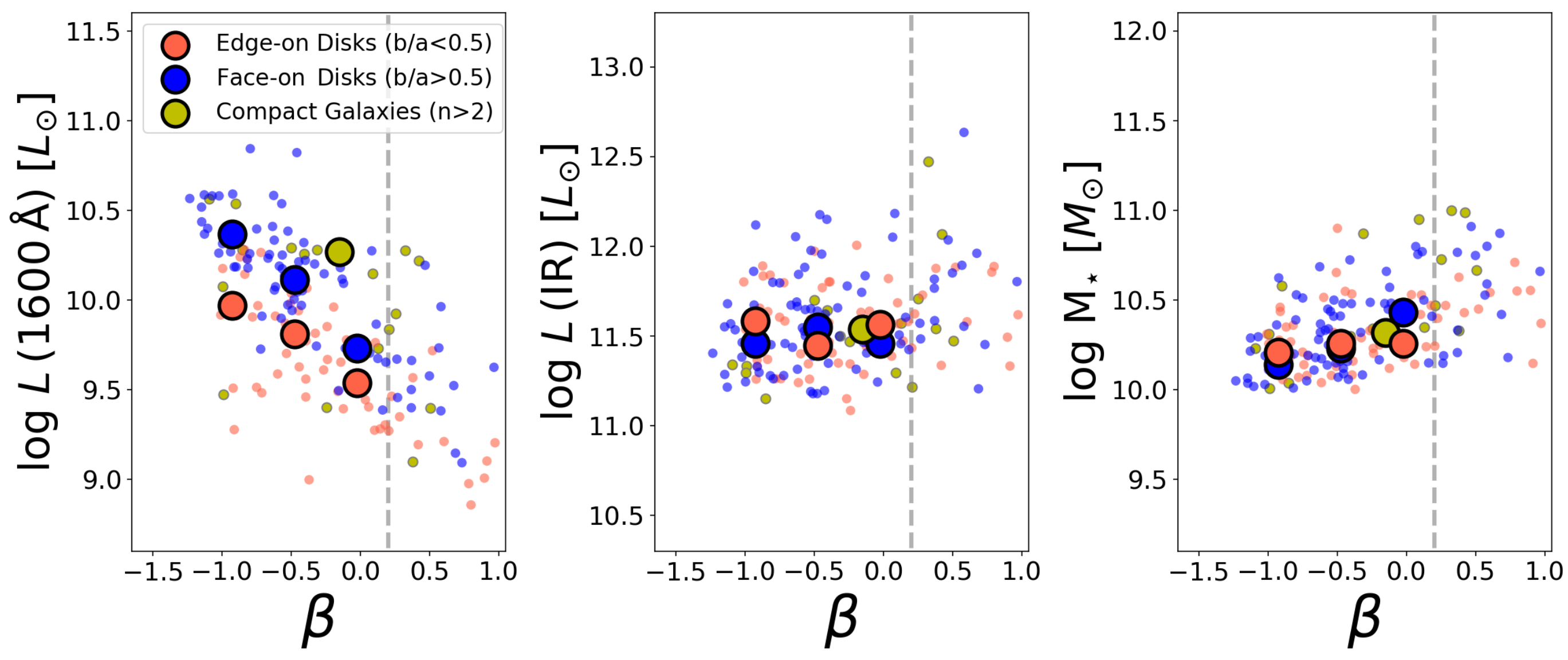}
	\caption{For the high-mass sample, edge-on galaxies have higher IRX (=$L$(IR)/$L$(1600\,\AA)) than face-on galaxies at the same $\beta$. This is mainly because the UV luminosities, $L$(1600\,\AA), of edge-on galaxies are lower than that those of face-on galaxies (left panel). At the same $\beta$, the infrared luminosities (middle panel) and stellar masses (right panel) are independent of inclination. \label{fig:irx_beta_samplecomp}}
\end{figure*}

\subsection{The High-Mass Sample $(10.0<\log\,M_\star/M_\sun<11.0)$}
\label{sec:rst_irxbeta_rst}

The IRX--$\beta$ relation for the high-mass sample is shown in Figure \ref{fig:irx_beta}.  As discussed in \S\ref{sec:measurement_incli}, the sample is divided into three populations: face-on disks ($b/a>0.5$, 69 objects), edge-on disks ($b/a<0.5$, 49 objects), and compact galaxies (11 objects). Compact galaxies are defined as those with H-band S\'ersic indexes greater than 2.0, and most of them have high central stellar mass surface densities (\citealt{Barro2017}; Appendix \ref{sec:appendix_morp}).

The major finding from Figure \ref{fig:irx_beta} is that face-on and edge-on galaxies exhibit clear offsets with respect to each other. At fixed $\beta$, edge-on galaxies have IRX values which are $\sim\,$0.3 dex greater on average than those of face-on galaxies. Correspondingly, the dust attenuation $A_\mathrm{UV}$ of edge-on galaxies is on average 0.9 mag higher than that of face-on galaxies at fixed $\beta$. Edge-on galaxies are around 0.1 dex above the \citet{Meurer1999} relation, whereas the face-on galaxies are around 0.2 dex below the relation. The locations of face-on galaxies overlap with the \citet{Heinis2013} sample of $z\sim 1.5$ star-forming galaxies. The discrepancy between this work and \citet{Heinis2013} can be due to the different U-band depths. The \citet{Heinis2013} work has shallower U-band depth (26.4 mag, shown as the gray dashed line in Figure \ref{fig:UVslopes}, compared to 27.9 mag for this work), which may miss galaxies with faint rest-frame FUV luminosities (and therefore high IRX values) at a given $\beta$. There is also difference regarding methodology. \cite{Heinis2013} stack galaxies but this work studies individual objects.
\par 
All galaxies in Figure \ref{fig:irx_beta} lie to higher IRX than the dust law of the Small Magellanic Cloud (\citealt{Gordon2003}), in contrast to studies at lower mass or higher redshift which favor the SMC relation (e.g., \citealt{Baker2001,Siana2009, Reddy2010, Capak2015, Watson2015,Bouwens2016,Koprowski2016, Pope2017, Reddy2017}).
\par 

\begin{figure*}
	\centering
	\includegraphics[width=7 in]{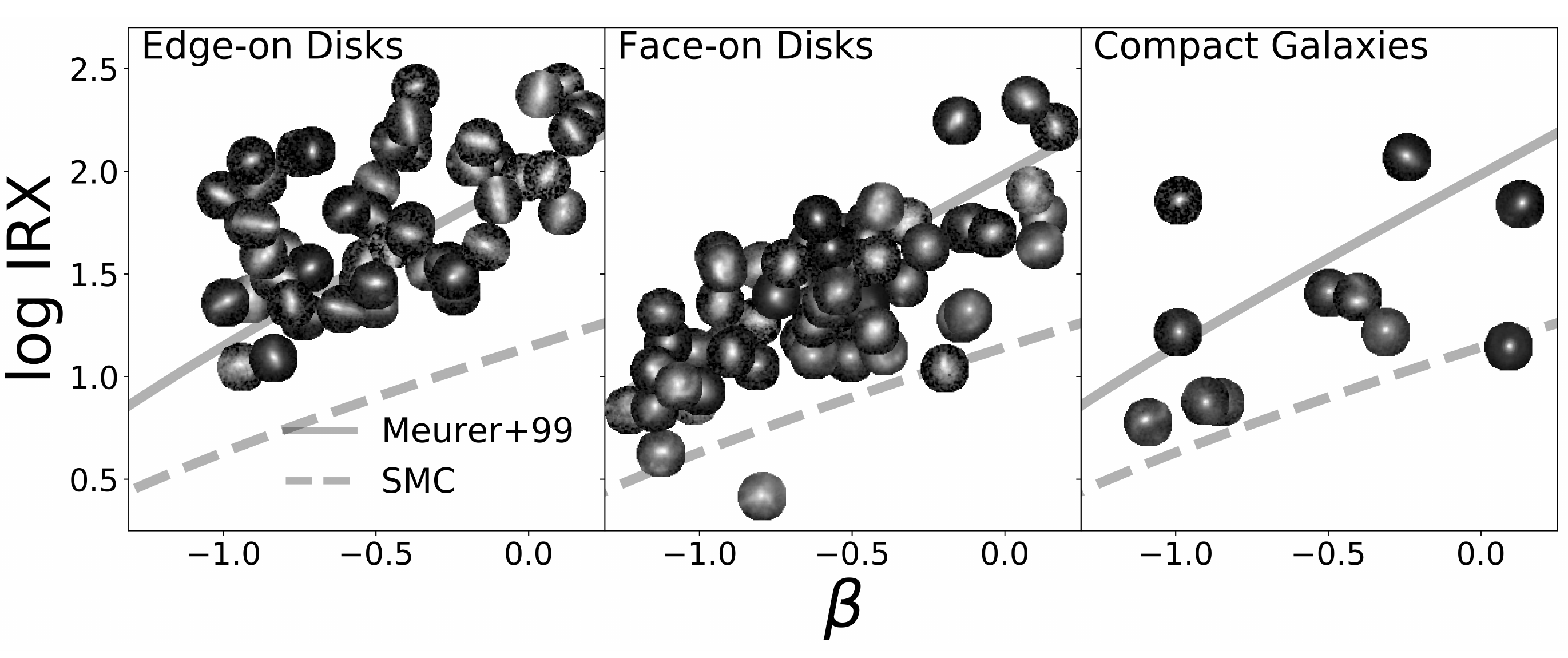}
	\caption{\emph{HST} H-band images of edge-on, face-on, and compact galaxies in the high-mass sample  ($10.0<\log\,M_\star/M_\sun<11.0$) are plotted on the IRX--$\beta$ diagram. Face-on and edge-on galaxies have extended disk features, while compact galaxies are small and featureless. Each image on the plot has a diameter of 2.4\arcsec\,(20 kpc at $z=1.5$). These images are taken from the public 3D-HST dataset \citep{Skelton2014}. \label{fig:postages}}
\end{figure*}

Properties of these disk galaxies are further compared in Figure \ref{fig:irx_beta_samplecomp}. In general, face-on galaxies and edge-on galaxies have the same average $L$(IR) and stellar masses, but their UV luminosities $L$(1600\,\AA) differ. This is expected because mid-IR to far-IR radiation is almost isotropic with little dust obscuration and the stellar masses derived from SED-fitting are not found to contain significant inclination-dependent errors \citep{Bell2001,Maller2009,Devour2017,Leslie2018a}. On the other hand, the far-UV radiation is obscured by dust inside galaxy disks and therefore viewing-angle dependent (e.g., \citealt{Jonsson2010,Chevallard2013,Leslie2018a}). At fixed $\beta$, edge-on galaxies have lower $L$(1600\,\AA) but the same $L$(IR) compared to face-on galaxies. Therefore, it is the difference in $L$(1600\,\AA) that leads to the difference in their IRX values (=$L$(IR)/$L$(1600\,\AA)) at fixed $\beta$.
\par 

As for compact galaxies, they generally have smaller sizes, rounder shapes ($b/a>0.5$), and are relatively featureless compared to disk galaxies (Figure \ref{fig:postages}). The compact objects seem to be on average located 0.2 dex below the locus of face-on disks on the IRX--$\beta$ diagram (Figure \ref{fig:irx_beta}). We do not discuss the physical interpretation of such a trend, due to the small sample size (11 objects) and large scatter on the IRX--$\beta$ diagram. A more comprehensive study will be possible in the future with a larger sample of compact galaxies.

\begin{figure*}
	\centering
	\includegraphics[width=7.0in]{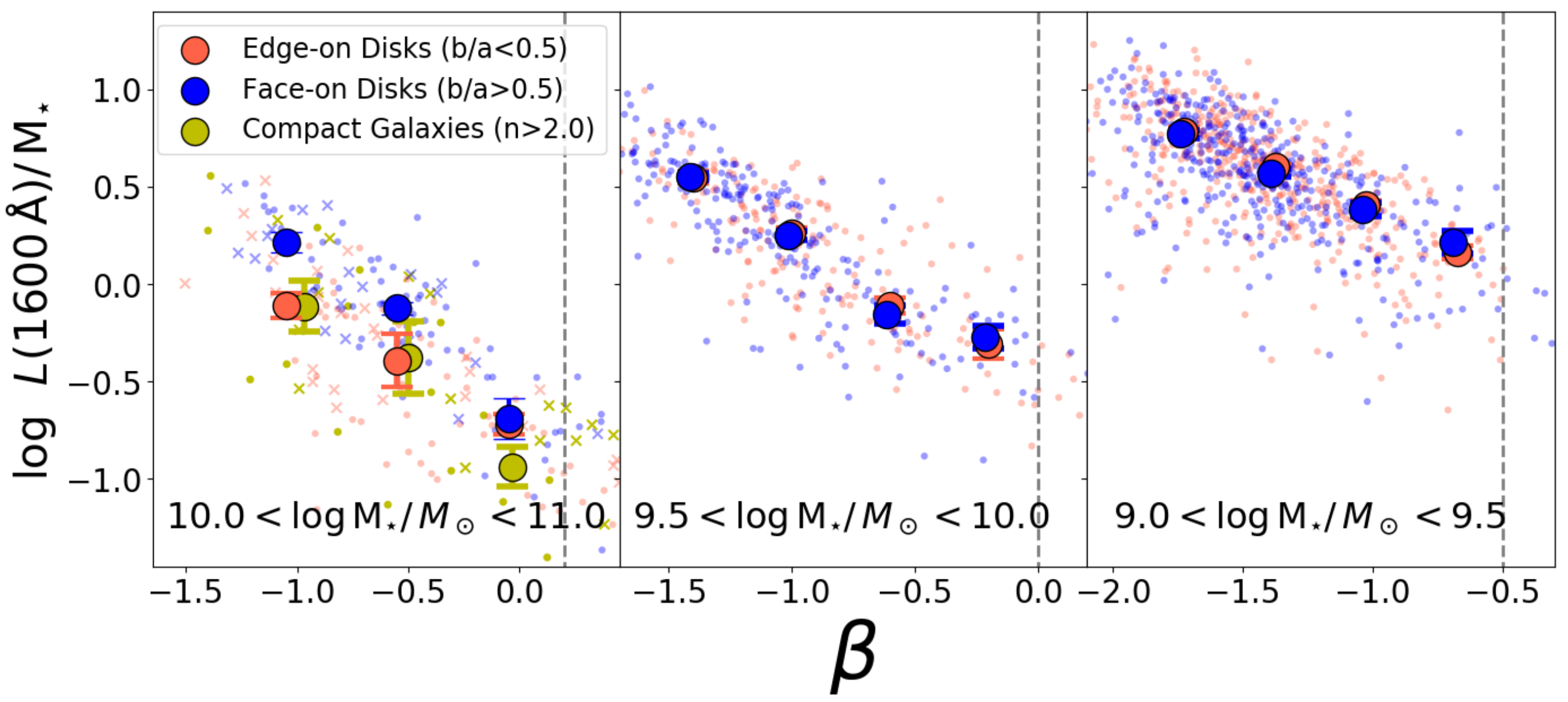}
	\caption{The  UV luminosity versus $\beta$ relation of high-mass disk galaxies varies with galaxy axis ratio (i.e., inclination, left panel), whereas that of galaxies in the low-mass sample does not (middle and right panels). This trend can be explained by that $A_\mathrm{UV}$ varies with axis ratio for high-mass disks at fixed $\beta$, which is found to be the case in Figure \ref{fig:irx_beta}, whereas it does not vary for galaxies in the low-mass sample. For the high-mass sample, we include galaxies both with and without reliable MIPS/24\,\micron\, fluxes (dots and crosses, respectively). Edge-on disks, face-on disks, and compact galaxies are indicated as red, blue, and yellow symbols on the left panel. For the low-mass sample on the middle and right panels, blue and red points indicate galaxies with low and high axis ratios ($b/a<0.5$ and $b/a>0.5$). The intrinsic shapes of these galaxies are discussed separately in \S \ref{sec:discussion_lowmass}. The UV luminosity $L$(1600\,\AA) is normalized by galaxy stellar mass, leading to a unit of $L_\sun\cdot M_\sun^{-1}$. Median trends are shown as large circles. Refer to \S\ref{sec:meurerlawdiscuss} and \S \ref{sec:discussion_lowmass} for details.
		\label{fig:umagvsb/a} }
\end{figure*}

In summary, we find that the IRX--$\beta$ relation is inclination-dependent for the massive star-forming galaxies that have MIPS 24\,\micron\ fluxes. Edge-on have higher IRX values than face-on galaxies at the same $\beta$. It means that edge-on galaxies have higher dust attenuation $A_\mathrm{UV}$ than face-on galaxies at the same $\beta$. Compact galaxies stay 0.2 dex lower than face-on galaxies on the IRX--$\beta$ diagram.
\subsubsection{Star-Forming Galaxies without 24\,\micron\ at $10.0<\log\,M_\star/M_\sun<11.0$}
\label{sec:rst_irxbeta_rst_wo24}

In the previous sample selection section (\S \ref{sec:selection_cuts}), 86 out of 219 massive galaxies at $10.0<\log\,M_\star/M_\sun<11.0$ are removed from the high-mass sample, because their 24\,\micron\ fluxes are not available due to the sensitivity and spatial resolution of \emph{Spitzer}/MIPS. The IRX (=$L$(IR)/$L$(UV)) versus $\beta$ relation of the 86 galaxies cannot be measured, because $L$(IR) has to be inferred from 24\,\micron\ fluxes. 

This section discusses whether the removal of these 86 galaxies biases our sample and influences the conclusion that edge-on galaxies have higher dust attenuation $A_\mathrm{UV}$ than face-on galaxies at the same $\beta$. An approach alternative to the IRX--$\beta$ diagram is applied to a sample which contains both the removed star-forming galaxies and the original high-mass sample. The approach assumes that the intrinsic (dust-free) UV luminosities (normalized by stellar mass) of face-on and edge-on galaxies at a given $\beta$ are the same, and examines whether their observed UV luminosities (measured at rest-frame 1600\,\AA) depend on inclination. If so, the difference in the observed UV luminosities must be caused by the difference in dust attenuation, which means that face-on galaxies and edge-on galaxies have different $A_\mathrm{UV}$ values at a given $\beta$. 

A total of 85 from the 86 removed galaxies are in the star-forming region of the UVJ diagram, and 59 of them are disk galaxies (i.e., S\'ersic index smaller than 2.0). These 59 galaxies are combined with disk galaxies in the high-mass sample, and the mass-normalized UV luminosity versus $\beta$ relation is shown on the left panel of Figure \ref{fig:umagvsb/a}.  The luminosity $L$(1600\,\AA) is normalized by stellar mass simply to help compare galaxies in different mass bins. The normalization does not change the relative trend between galaxies with different inclinations or axis ratios, because we find that their stellar masses are on average the same at a given $\beta$.

The major finding of Figure \ref{fig:umagvsb/a} (left panel) is that at fixed $\beta$, the UV luminosities of edge-on galaxies are approximately 0.3 dex lower than those of face-on galaxies. Assuming that the intrinsic UV luminosities are the same, this means that at fixed $\beta$, edge-on galaxies have higher dust attenuation, $A_\mathrm{UV}$, than face-on galaxies. This is the same as what we find for the 24\,\micron-detected galaxies in  Figure \ref{fig:irx_beta_samplecomp}. 

Finally, we validate the assumption that the \emph{intrinsic} UV luminosities of face-on and edge-on galaxies at a given $\beta$ are on average the same. The intrinsic UV luminosity is proportional to SFR for star-forming galaxies \citep{Kennicutt2012, Madau2014a}. So if the intrinsic luminosities of face-on and edge-on galaxies at a given $\beta$ differ and $A_\mathrm{UV}$ values are the same, then for all the edge-on galaxies we study in this section,  their SFR would be $\sim$0.3 dex smaller than those of the face-on galaxies on average. This is not expected because physical properties of disk galaxies do not depend on the viewing angle. Two more findings support this argument. First, we examined the SFR of these galaxies using the SED-fitting values by \cite{Fang2018} and find no difference between face-on galaxies and edge-on galaxies at the same $\beta$. Second, the stellar masses of face-on and edge-on galaxies at the same $\beta$ are found to be the same. If their average SFR values differ, the distributions of face-on and edge-on galaxies on the SFR--$M_\star$ diagram will be different, which is not found by studies of the SFMS relation (e.g., \citealt{Speagle2014})

In summary, we conclude that the $A_\mathrm{UV}$--$\beta$ relation is inclination-dependent for star-forming galaxies at $10.0<\log\,M_\star/M_\sun<11.0$, even including galaxies without MIPS/24\,\micron\ detection. Edge-on galaxies have lower observed UV luminosities ($\sim$\,0.3 dex) than face-on galaxies at a given $\beta$. It means that edge-on galaxies have higher dust attenuation $A_\mathrm{UV}$ than face-on galaxies at the same $\beta$.

\begin{figure*}
	\centering
	\includegraphics[width=7in]{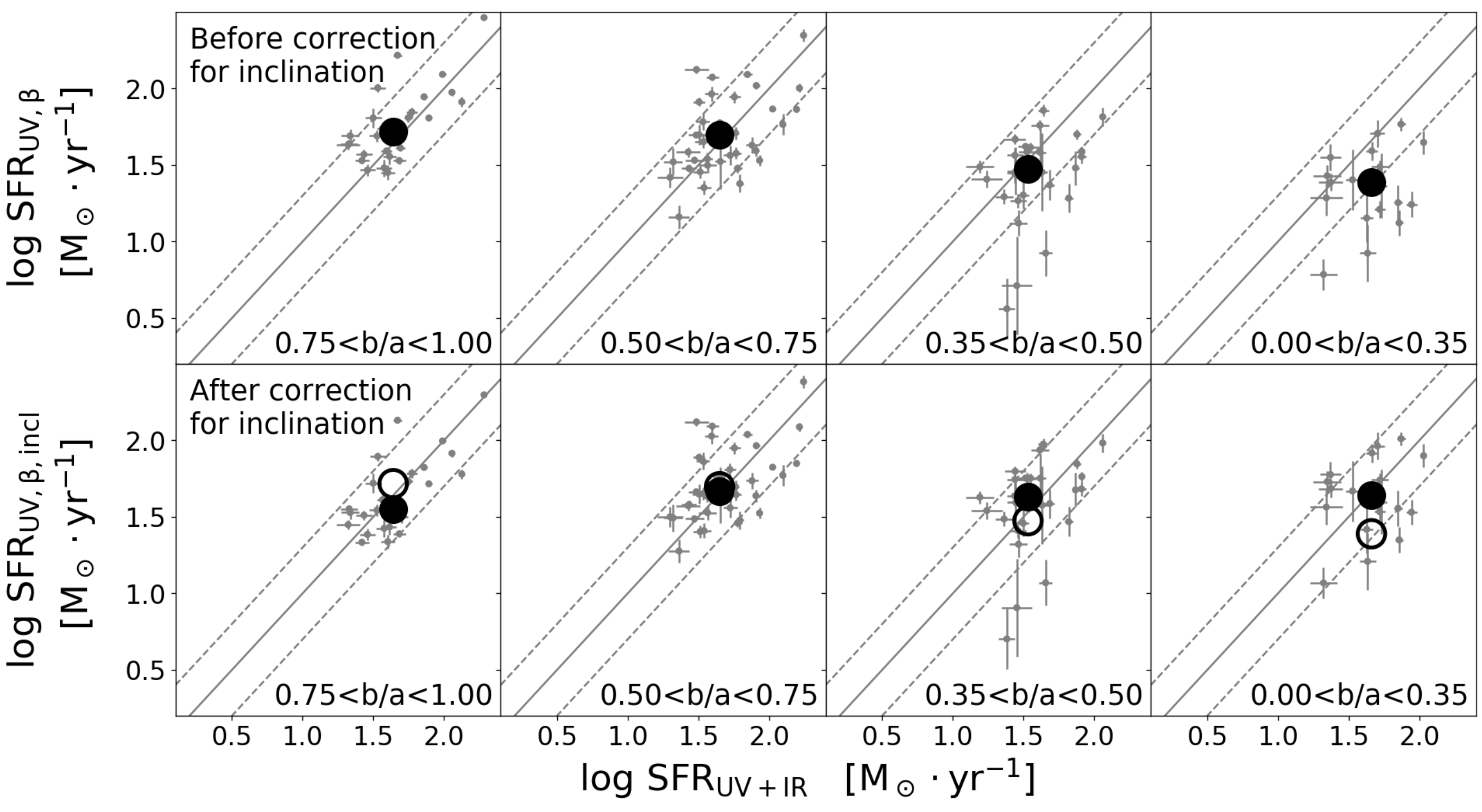}
	\caption{\label{fig:sfr_cali} The UV-based SFR values  before and after the inclination correction to the $A_{\mathrm{UV}}-\beta$ relation (SFR$_{\mathrm{UV,\;\beta}}$ and SFR$_{\mathrm{UV,\;\beta, incl}}$) are compared with SFR$_{\mathrm{UV+IR}}$ in the top and bottom panels, respectively. Disk galaxies in the high-mass sample are separated into four axis ratio bins, where large (small) axis ratio indicates face-on (edge-on) systems. The correction has the most impact on galaxies with high inclinations ($b/a<0.35$), and brings UV-based SFR closer to SFR$_\mathrm{UV+IR}$. 
	Small dots indicate individual galaxies, and large filled circles are median values for each panel.  For each axis ratio bin, the large open circle on the bottom panel marks the same location as for the filled circle on the top panel. Errorbars in the plot indicate photometric uncertainties. Dashed lines indicate a 0.3 dex scatter from the one-to-one relation. }
\end{figure*}

\subsection{The Low-Mass Sample $(9.0<\log\,M_\star/M_\sun<10.0)$}
\label{sec:meurerlawdiscuss}
Following \S \ref{sec:rst_irxbeta_rst_wo24}, we keep using the UV luminosity versus UV slope diagram to examine whether the $A_\mathrm{UV}$--$\beta$ relation depends on the axis ratio for the low-mass sample. Most of these galaxies do not have 24\,\micron\ fluxes to derive the IRX--$\beta$ relation, due to the sensitivity of MIPS/24\,\micron. The previous section (\S \ref{sec:rst_irxbeta_rst_wo24}) shows that if galaxies with different axis ratios have different observed UV luminosities at a given $\beta$, then such difference indicates that they have different dust attenuation values $A_\mathrm{UV}$. The following discussions about the low-mass sample make use of the same argument.
\par 

The $L(1600\,\mbox{\AA})/\mathrm{M}_\mathrm{\star}-\beta$ relations of low-mass galaxies are shown on the middle and right panels of Figure \ref{fig:umagvsb/a}. They do not vary with galaxy axis ratio in the ranges $9.0<\log\,M_{\star}/M_{\sun}<9.5$ and $9.5<\log\,M_{\star}/M_{\sun}<10.0$. This is even the case for the galaxies with $\beta$ as red as those of the high-mass galaxies.  At a given $\beta$, the same observed UV luminosities of galaxies with high axis ratios ($b/a>0.5$) and low axis ratios ($b/a<0.5$) indicate that these galaxies may have the same UV dust attenuation. 

In summary, it is likely that galaxies in the low-mass sample do not have an $A_\mathrm{UV}$--$\beta$ relation that varies with the axis ratio. Possible physical reasons for this are discussed in \S\ref{sec:RT_mode}.

\section{High-mass sample: An Inclination-dependent Calibration of $A_{\mathrm{UV}}$ and UV-based SFR}
\label{sec:rst_sfr}
This section focuses on the high-mass sample ($10.0<\log\,M_\star/M_\sun<11.0$), and shows that the inclination-dependence of the IRX--$\beta$ relation leads to a discrepancy between two commonly used SFR indicators, $\mathrm{SFR}_{\mathrm{UV+IR}}$ and $\mathrm{SFR}_{\mathrm{UV,\,\beta}}$ (\S \ref{sec:sfr_compare}). The definitions of $\mathrm{SFR}_{\mathrm{UV+IR}}$ and $\mathrm{SFR}_{\mathrm{UV,\,\beta}}$ are described in \S \ref{sec:measurement_sfr}.
To remove this discrepancy, we compute corrections to the $A_\mathrm{UV,\,Meurer}$--$\beta$ relation (Equation \ref{eqn:meurerrelation}) and the associated equation measuring $\mathrm{SFR}_{\mathrm{UV,\,\beta}}$ in \S\ref{sec:sfr_cali}.

\subsection{Discrepancy Between $\mathrm{SFR}_{\mathrm{UV+IR}}$ and $\mathrm{SFR}_{\mathrm{UV,\,\beta}}$}
\label{sec:sfr_compare}
The discrepancy between $\mathrm{SFR}_{\mathrm{UV+IR}}$ and $\mathrm{SFR}_{\mathrm{UV,\,\beta}}$ as a function of galaxy inclination is shown in four ranges of axis ratio in Figure \ref{fig:sfr_cali}. For the most face-on galaxies ($b/a>0.75$), $\mathrm{SFR}_{\mathrm{UV,\,\beta}}$ is on average 0.07 dex greater than $\mathrm{SFR}_{\mathrm{UV+IR}}$, and for the most edge-on galaxies ($b/a<0.35$) it is 0.25 dex smaller. 
\par 
Such a discrepancy is caused by the dust attenuation $A_\mathrm{UV}$ being overestimated/underestimated for face-on/edge-on galaxies when deriving $\mathrm{SFR}_{\mathrm{UV,\,\beta}}$ with the $A_\mathrm{UV}-\beta$ relation by \cite{Meurer1999}. As already shown in Figure \ref{fig:irx_beta}, for the high-mass sample at $z\sim 1.5$, the local starburst relation of \citet{Meurer1999} overestimates IRX and $A_\mathrm{UV}$ for face-on galaxies, but underestimates them for edge-on galaxies. $\mathrm{SFR}_{\mathrm{UV+IR}}$ does not contain such systematic error \citep{Leslie2018}. As a result, $\mathrm{SFR}_{\mathrm{UV,\,\beta}}$ is on average larger than $\mathrm{SFR}_{\mathrm{UV+IR}}$ for face-on galaxies and smaller for edge-on galaxies. 

\subsection{Inclination-Dependence of $A_{\mathrm{UV}}$ and SFR$_\mathrm{UV}$}
\label{sec:sfr_cali}
\begin{figure}
	\centering
	\includegraphics[width=3.2 in]{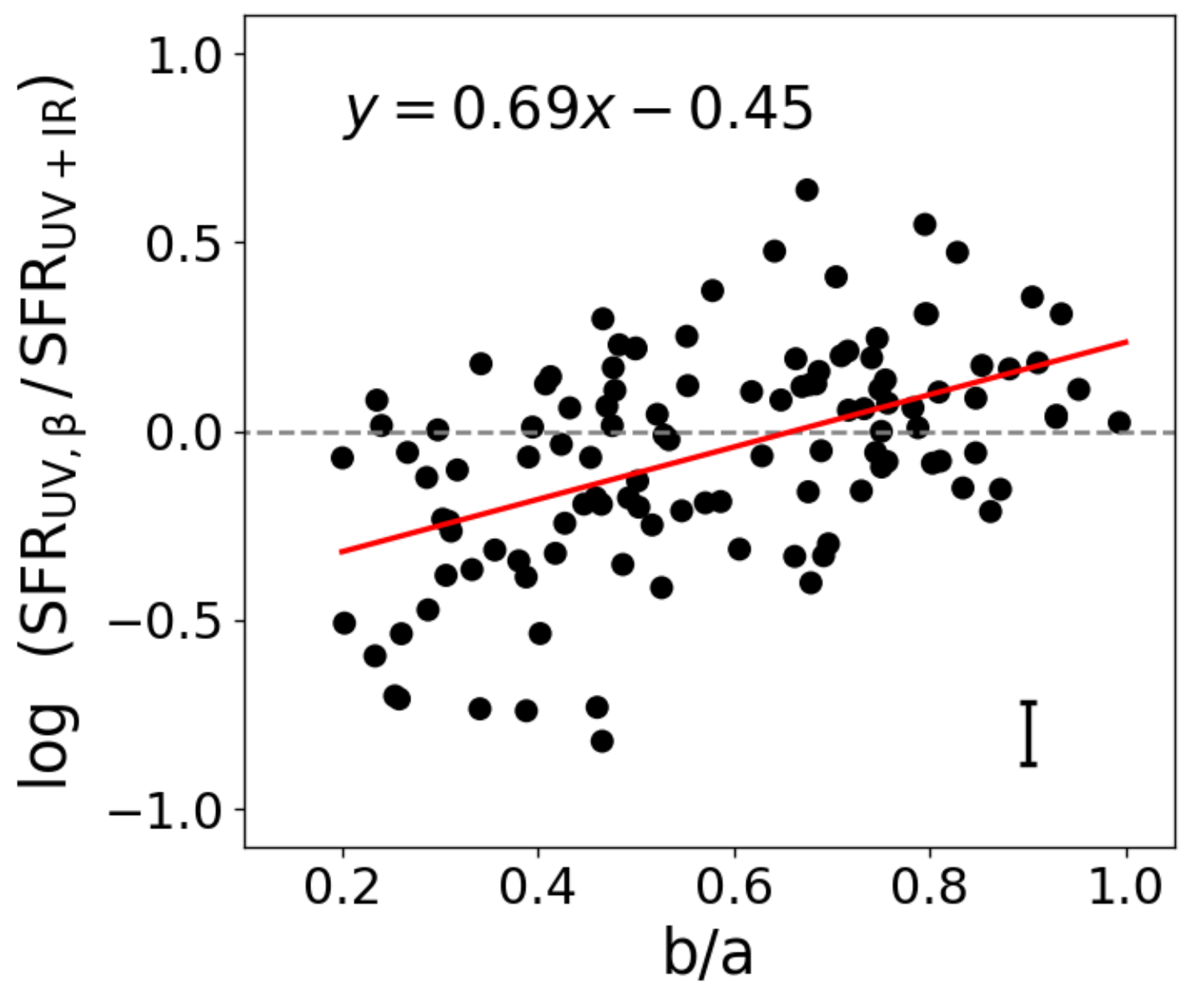}
	\caption{\label{fig:sfr_fit} For the high-mass disk galaxies, the SFR values inferred from UV spectral slopes (Equation \ref{eqn:sfrbeta}) are generally smaller than SFR$_\mathrm{UV+IR}$ (Equation \ref{eqn:sfr_uvir}) at high inclination and greater at low inclination. The median value of SFR$_\mathrm{UV, \beta}$/SFR$_\mathrm{UV+IR}$ for all the galaxies is 0.95. The red line is a linear fit to the individual points. It is used to remove inclination-dependence for the new $A_\mathrm{UV}$ and SFR indicators in Equations \ref{eqn:meurer_new} and \ref{eqn:sfrbeta_new}. The errorbar indicates the typical uncertainty of individual galaxies.}
\end{figure}

In this section, an inclination correction to the $A_{\mathrm{UV,\,Meurer}}$--$\beta$ relation is derived for high-mass galaxies. To do this, a linear relation is fit to the quantity $\log\,(\mathrm{SFR}_{\mathrm{UV,\,\beta}}\,/\,\mathrm{SFR}_{\mathrm{UV+IR}})$ versus the axis ratio of disk galaxies, as shown in Figure  \ref{fig:sfr_fit}. This relation accounts for the part of dust attenuation correction that is relevant to the inclination of galaxies to the line-of-sight. Then, the following equation of UV dust attenuation is derived from the fit:
\begin{equation}
 A_\mathrm{UV,\,incl}(\beta,\ b/a)=A_\mathrm{UV,\,Meurer}(\beta)-1.73\,(b/a-0.67)\  \mathrm{[mag]}, \label{eqn:meurer_new}
\end{equation}
where $A_\mathrm{UV,\,Meurer}(\beta)$ is from the original \citet{Meurer1999} relation (Equation \ref{eqn:meurerrelation}), and $A_\mathrm{UV,\,incl}$ is the new indicator of UV attenuation. Correspondingly, the formula for UV-based SFR is updated by replacing $A_{\mathrm{UV,\,Meurer}}$ with $A_{\mathrm{UV,\,incl}}$ in Equation \ref{eqn:sfrbeta}:
\begin{equation}
\mathrm{SFR}_{\mathrm{UV,\,\beta,\, incl}}=1.72\times 10^{-10}L(1600\,\mbox{\AA})\cdot 10^{0.4\,A_{\mathrm{UV,incl}}}. \label{eqn:sfrbeta_new}
\end{equation}

These formulae apply to star-forming galaxies at $z\sim 1.5$ and $10.0<\log\,M_\star/M_\sun<11.0$. 
For an edge-on galaxy with $b/a=0.25$, the inclination correction term results an increase of 0.3 dex from $\mathrm{SFR}_{\mathrm{UV,\,\beta}}$ to $\mathrm{SFR}_{\mathrm{UV,\,\beta,\,incl}}$. This is comparable to the  SFMS scatter ($\sim\,$0.3 dex, \citealt{Speagle2014}). Studies of the SFMS at similar redshifts as our sample should use this correction, if they rely on SFRs derived from UV luminosities and UV spectral slopes.
\par
The inclination-corrected star-formation rate, i.e., $\mathrm{SFR}_{\mathrm{UV,\,\beta,\, incl}}$, is compared with $\mathrm{SFR}_{\mathrm{UV+IR}}$ in the lower panels of Figure \ref{fig:sfr_cali}.  Compared with top panels, the correction has the most impact on galaxies with high inclinations ($b/a<0.35$) and brings UV-based SFR closer to SFR$_\mathrm{UV+IR}$. For the entire sample of high-mass disk galaxies, the standard deviation of $\log\,\mathrm{SFR}_{\mathrm{UV,\,\beta,\, incl}}-\log\,\mathrm{SFR}_{\mathrm{UV+IR}}$ is 0.25 dex, slightly smaller than that of $\log\,\mathrm{SFR}_{\mathrm{UV,\,\beta}}-\log\,\mathrm{SFR}_{\mathrm{UV+IR}}$, which is 0.29 dex.
\par 
This correction is consistent with recent work on ``local'' ($z\sim 0.07$) disk galaxies by \cite{Leslie2018}. For galaxies with stellar masses above $10^{10.2}\,M_\sun$, they find that edge-on disks have higher $A_\mathrm{UV}$ than face-on disks at the same $\beta$, which is also indicated by Equation \ref{eqn:meurer_new}. \cite{Leslie2018} also study $z\sim 0.7$ galaxies with similar masses, above $10^{10.2}\,M_\sun$, and do not find an inclination-dependence. However, it is possible that the most dusty edge-on galaxies may be missing from their sample due to \emph{GALEX} sensitivity (25.5 mag in the near-UV band).

\section{Implications for the Shapes of Dust Attenuation Curves}
\label{sec:rst_irxbeta_interp}
From the inclination-dependence of the IRX--$\beta$ relation, we infer that at fixed UV dust attenuation $A_\mathrm{UV}$, massive edge-on galaxies have flatter (grayer) UV dust attenuation curves than face-on galaxies. This comes from a comparison of the UV spectral slopes for galaxies at the same IRX value.

\subsection{The relation between the IRX--$\beta$ relation and dust attenuation curve shapes}
For the high-mass sample, Figure \ref{fig:irx_beta} indicates that edge-on galaxies have on average lower values of $\beta$ than the face-on galaxies with the same IRX.  IRX is a robust measure of dust attenuation at rest-frame 1600\,\AA, $A_{\mathrm{UV}}$, for star-forming galaxies with a broad range of dust geometry and dust grain types \citep{Gordon2000, Witt2000, Overzier2011, Buat2012, Narayanan2018}. Therefore, the IRX--$\beta$ relations indicate that edge-on galaxies lie to lower $\beta$ than face-on galaxies at the same $A_{\mathrm{UV}}$. 

This difference in $\beta$ values can be explained if the UV attenuation curves of edge-on galaxies are flatter (i.e., grayer) than those of face-on galaxies\footnote{In this work, dust attenuation curves are normalized in the rest-frame far-UV before before their shapes are compared with each other.}, at a given $A_\mathrm{UV}$. With a flatter attenuation curve, there is a smaller difference between far-UV and near-UV attenuation. Therefore, at a given $A_\mathrm{UV}$, such galaxies are less reddened in the UV and have bluer $\beta$'s than those with steeper attenuation curves.

We also fit the dust attenuation curve shapes with a widely used function form by \cite{Noll2009}, in order to allow future relevant works to make a quantitative comparison with ours. The function form uses two parameters, $\delta$ and $E_b$, to quantify the slope and the 2175\AA\ bump amplitude of a dust attenuation curve, relative to the Calzetti dust law. Greater $\delta$ indicates a flatter (grayer) attenuation curve. The attenuation curves are drawn from the radiative transfer simulations that match observed galaxies on the IRX--$\beta$ diagram (see Section \ref{sec:RT_mode}). At $\beta=-0.6$, for example, the dust attenuation curve that matches edge-on galaxies has $\delta=0.22$, and the curve for face-on galaxies has $\delta=0.12$. In other words, edge-on galaxies are fit with flatter (grayer) attenuation curves than face-on galaxies. This is in agreement with Fig.\ 17 of \cite{Buat2012}, which shows that a greater $\delta$ value places galaxies upward on the IRX--$\beta$ diagram (also refer to Salim et al.\ 2018, submitted).
\par 

Finally, we consider whether the different $\beta$ values at the same IRX can be caused by a difference in galaxy stellar populations rather than different dust attenuation curves. In this scenario, at the same IRX, the UV spectral slopes of galaxies are reddened by the same amount due to dust, but the \emph{intrinsic} dust-free UV spectral slopes ($\beta_0$) of face-on galaxies are redder than those of edge-on galaxies. This explanation is not favored because face-on and edge-on galaxies in the high-mass sample are in the same mass and redshift range (Figure \ref{fig:irx_beta_samplecomp}), and the underlying stellar populations of galaxies should not vary with inclination at the depth of photometry used in this work. As one more test to show that variation of $\beta_{0}$ does not fully explain the observed difference in $\beta$, we take previous stellar population modeling results, i.e., Fig.\ 3 of \citet{Grasha2013}, and examine $\beta_0$ as a function of stellar metallicity and stellar age. If assuming a constant star-formation history, a difference of 0.4 in $\beta_{0}$, which is the amount required to explain the inclination dependence, corresponds to a difference of approximately 0.6 Gyr in stellar age, or changing the stellar metallicity from $Z=0.0004$ to $Z=0.02$. Such difference in stellar age or stellar metallicity is not expected between face-on and edge-on star-forming galaxies at a given mass range.

As for the low-mass sample, \S \ref{sec:meurerlawdiscuss} shows that its UV luminosity versus $\beta$ relation does not have a dependence on axis ratio,  and therefore the $A_\mathrm{UV}$--$\beta$ relation likely does not either. This indicates that the dust attenuation curves of low-mass galaxies likely do not vary with axis ratio at fixed $A_\mathrm{UV}$.

\subsection{Comparison with Other Studies}
This section compares with other studies of dust attenuation curve shapes as a function of inclination, in the mass range $10.0<\log\,M_\star/M_\sun<11.0$. 
	
These studies find that dust attenuation curves become flatter (grayer) as galaxy inclination increases for local galaxies (e.g., \citealt{Wild2011, Chevallard2013, Battisti2017,Salim2018}), which are consistent with our conclusion.
	
Interestingly, such inclination trend may be attributed to the difference in $A_V$ \emph{to the first order} (e.g., \citealt{Salmon2016}; Salim et al.\ 2018, submitted). This is because edge-on galaxies on average have higher $A_V$ than face-on galaxies and the dust attenuation curve gets flatter (grayer) as $A_V$ increases \citep{Witt2000, Pierini2004, Inoue2005, Chevallard2013, Salmon2016, Leja2017, Narayanan2018b}. This paper does not directly study the trend of dust attenuation curve shape with $A_V$, but studies galaxies with various inclinations at a certain $A_\mathrm{UV}$ value. We note that there is no strictly one-to-one correspondence between $A_\mathrm{UV}$ and $A_V$ (e.g., \citealt{Salim2018}). Nevertheless, for our future work, it is an interesting topic to explore the role of $A_V$  in the IRX--$\beta$ relation beyond the local universe.

\begin{figure*}
	\centering
	\includegraphics[width=6.5in]{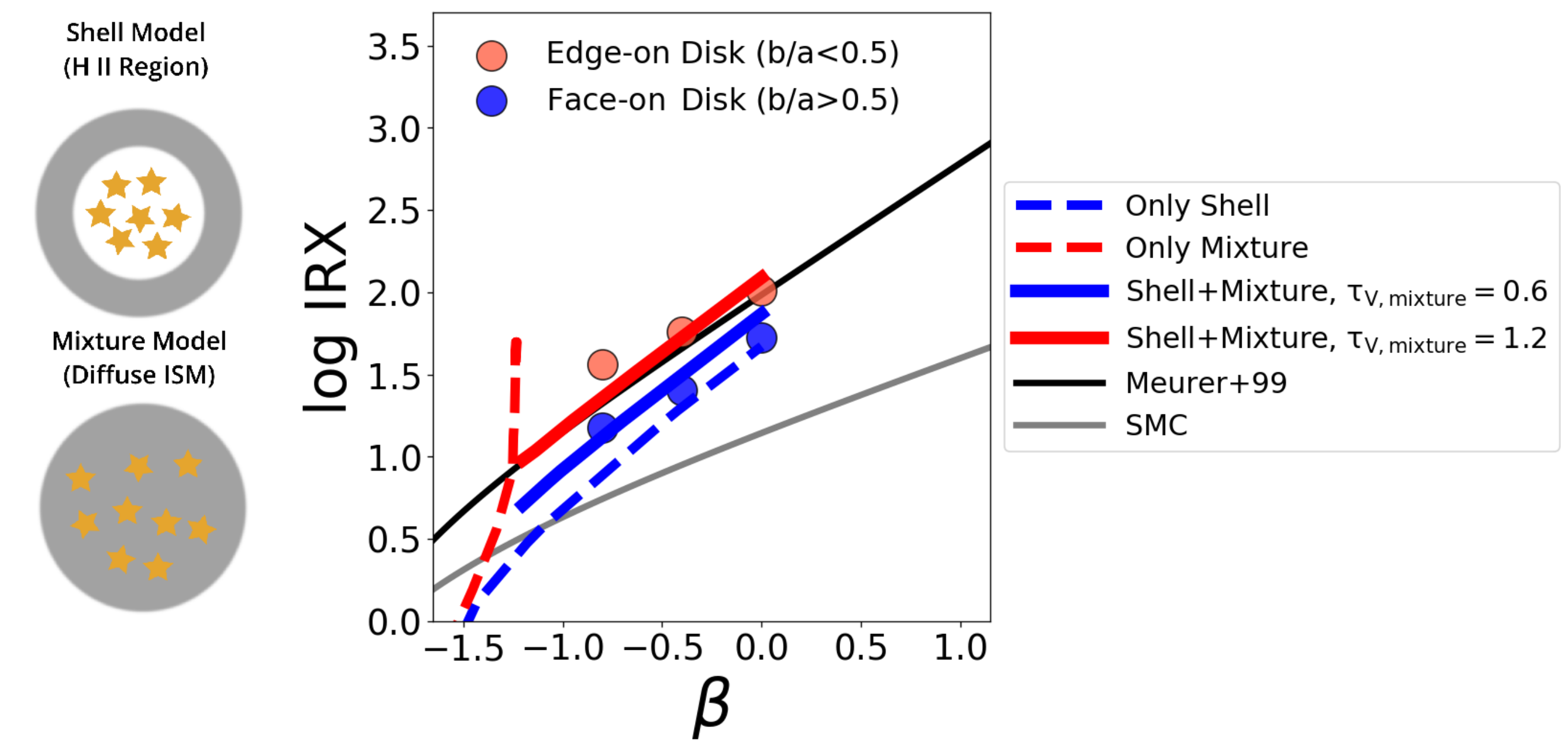}
	\caption{The observed inclination-dependent IRX--$\beta$ relation can be explained by a physical model using radiative transfer results \citep{Witt2000,Seon2016,Law2018}. Two-component dust models (solid curves), which combine the two illustrated types of dust-star distribution, match the observation on the IRX--$\beta$ diagram. In such models, edge-on galaxies have higher dust optical depth from the diffuse interstellar medium, i.e., the shell model component, than face-on galaxies. 
	The models adopt solar stellar metallicity, a stellar age of 50 Myr, and homogeneous Milky Way-type dust grain. The dust attenuation curves from radiative transfer models have significant bump feature at 2175\AA. The UV slope $\beta$ is measured in the way consistent with the observational approach after the model spectra is redshifted to $z=1.5$. Colored points are the observed median positions of high-mass galaxies in Figure \ref{fig:irx_beta}.
		\label{fig:dirtygrid}}
\end{figure*}

\section{Star-Dust Geometry of High-Mass Galaxies}
\label{sec:RT_mode}
This section tries to explain the inclination-dependence of the IRX--$\beta$ relation using physical models of the dust and star distribution inside galaxies.

Such models come in two flavors: single-component and two-component. Single-component models assume that stars of all ages are distributed relative to dust in the same way. Two-component models
assume that young stars are subject to more dust attenuation than older stars (e.g., \citealt{Calzetti2001, Charlot2000}). Specifically, dust in the diffuse interstellar medium (ISM) of galaxies forms a uniform mixture with stars of all ages, while young O/B stars additionally reside in \ion{H}{2} regions surrounded by dust cocoons.

It is concluded that only a two-component dust model \citep{Calzetti1994,Charlot2000} can explain the offset in the IRX--$\beta$ diagram between face-on and edge-on galaxies that we find for the high-mass sample. 
\subsection{Single-Component Dust Models}
\label{sec:discuss_singlemodel}

Typical single-component dust models can be a uniform mixture of stars and dust, or a dust shell surrounding the stars (refer to Figure \ref{fig:dirtygrid}, left). For a given single-component model and a given dust grain type, the dust optical depth along the line of sight ($\tau_{\mathrm{UV}}=0.921 A_\mathrm{UV}$) is the only variable that controls the shape of dust attenuation curve \citep{Witt2000, Seon2016, Popping2017}. However, Figure \ref{fig:irx_beta} and \S \ref{sec:rst_irxbeta_interp} show that edge-on galaxies have grayer UV dust attenuation curves than face-on galaxies, even if they have the same $A_\mathrm{UV}$. This means that besides $A_\mathrm{UV}$, galaxy inclination influences the dust attenuation curve shape as a second independent variable. Single-component models cannot explain such observed inclination-dependence, since they only predict the same dust attenuation curve shape at a given $A_\mathrm{UV}$.  
 
\subsection{Two-Component Dust Models}
\label{sec:discuss_doublemodel}

Now we turn to two-component dust models.   
These models assume that UV photons are first attenuated by dust in a shell component and then by dust in a mixture component when traveling from the young stars inside galaxies to the observer (see Figure \ref{fig:dirtygrid}, left).
We show that edge-on galaxies are expected to have relatively higher dust optical depth from the mixture component than face-on galaxies, and this is the most probable cause of the inclination-dependence of the IRX--$\beta$ relation.

To illustrate this scenario, we use radiative transfer models to generate IRX--$\beta$ relations (Figure \ref{fig:dirtygrid}). The Milky-Way dust grain model by \citet{Witt2000} (their Table 1) and the updated radiative transfer results\footnote{\url{https://seoncafe.github.io/MoCafe.html}} by \citet{Seon2016} are adopted. Stellar population synthesis models are taken from \citet{Bruzual2003}. The shell model spans broad ranges in both IRX and $\beta$. As for the mixture model, the UV spectral slope stays at a small constant value as IRX increases, because the UV light in such a configuration is always dominated by the less attenuated stars near the  galaxy surface (see also \citealt{Popping2017}). However, neither of these two single-component models match the observation results.

Only two-component models recover the locations of observed galaxies on the IRX--$\beta$ diagram. For these models, we use $\tau_{\mathrm{V, mixture}}$ to represent the effective dust optical depth from the mixture component. A model with $\tau_{\mathrm{V, mixture}}=0.6$ matches face-on galaxies, and a model with a higher value of 1.2 matches edge-on galaxies. \textit{This indicates that the inclination-dependent IRX--$\beta$ relation is caused by that edge-on galaxies have a higher fraction of dust optical depth from the mixture component than face-on galaxies. }This is physically plausible, given that the path-length through the diffuse ISM increases with galaxy inclination, whereas the dust optical depth from around \ion{H}{2} regions does not change due to their near-spherical shell shape (\citealt{Tuffs2004, Yip2010, Popescu2010, Wild2011, Chevallard2013, Battisti2017}). 

 One additional effect may also contribute to the inclination-dependence of the IRX--$\beta$ relation. For realistic star-forming galaxies at $z\sim 1.5$, dust attenuation decreases with the radial distance to galactic centers \citep{Nelson2016, Liu2017,Wang2017,Tacchella2018}. Their less dusty outskirts, which are blue in the UV, cover most of the sight-lines when galaxies are viewed edge-on. This makes edge-on galaxies have even bluer observed UV spectral slopes than those predicted from the simple two-component dust models.

In summary, these models imply that edge-on galaxies have bluer UV spectral slopes ($\beta$) than face-on galaxies at a given IRX because of a higher contribution from the ``mixture component'' of star-dust distribution. Similar radiative transfer results are also found and analyzed in detail by \citet{Charlot2000}, \citet{Nordon2013}, and \citet{Popping2017}. 

\section{Star-Dust Geometry of Low-Mass Galaxies}
\label{sec:discussion_lowmass}
We do not find a dependence of galaxy axis ratio in the $L(1600\,\mbox{\AA})/\mathrm{M}_\mathrm{\star}-\beta$ relation at $9.0<\log\,M_\star/M_\sun<10.0$ in \S\ref{sec:meurerlawdiscuss}. This conclusion holds even if we only select galaxies with low S\'ersic indexes ($n<2.0$). The most probable reason is that low-mass galaxies are more prolate and/or dust is distributed highly irregularly in these galaxies. In such circumstances, the relative distribution of young stars and dust does not regularly vary with the projected axis ratio, and therefore neither does the IRX--$\beta$ relation. 

Studies of galaxy axis ratio distribution  in this mass and redshift range indicate that more than half of the star-forming galaxies have prolate shapes for their stellar component \citep{VanderWel2014, Zhang2018}. Recent simulations suggest that this is likely because the gravitational potential of these galaxies is driven by the prolate inner parts of dark matter halos \citep{Ceverino2015,Tomassetti2016}. The transition from oblate shape at high mass to prolate shape at low mass happens at around $10^{9.4\pm 0.4}\,M_\sun$  for $z>1$ in the simulation \citep{Tomassetti2016}, which matches the mass range we study ($9.0<\log\,M_\star/M_\sun<10.0$). These simulations further show that the distribution of gas in these low-mass galaxies, where dust resides, is highly amorphous.

Observationally, gas kinematic studies provide the most direct support of the suggested scenario about gas distribution inside these galaxies. Low-mass galaxies at around $z=1.5$ tend to be more often dominated by disordered motions than rotational motions  \citep{Law2009, Kassin2012, Simons2016, Simons2017}. This implies that in general disks are not in place and the distribution of gas and likely also dust have disturbed morphology at this epoch. 

Another possible reason for the absence of an inclination-dependence is that the UV spectral slopes of low-mass galaxies are driven by the stochastic variation of stellar population properties, including SFR and stellar metallicity (e.g., \citealt{Guo2016,Hopkins2017,Sparre2017,Faucher-Gigu2017}), which further changes the \emph{intrinsic} UV spectral slope. For $\beta<-1.4$, \citet{Reddy2017} shows that galaxies have bluer $\beta$ values at a lower $L_\mathrm{UV}$, but  IRX, as a direct probe of dust attenuation, does not vary with $L_\mathrm{UV}$ (see Figures 1 and 8, and \S 5.3 therein). This indicates that the change of observed $\beta$ values with $L_\mathrm{UV}$ is not due to the change of dust attenuation, but rather a change of \emph{intrinsic} UV spectral slopes. However, in our work, galaxies in the low-mass sample have spectral UV slopes as red as $-0.5$. Their UV slopes are less likely to be solely driven by the variation of intrinsic UV spectral slopes.

\section{Conclusions} \label{sec:conclude}
The dust attenuation properties of star-forming galaxies at $1.3<z<1.7$ are examined as a function of galaxy inclination in the GOODS-South and GOODS-North fields using data from the CANDELS survey \citep{Grogin2011, Koekemoer2011}. Two ranges of galaxy stellar mass are studied: high-mass ($10.0<\log\,M_{\star}/M_{\sun}<11.0$) and low-mass ($9.0<\log\,M_{\star}/M_{\sun}<10.0$). The IRX--$\beta$ diagram \citep{Meurer1999} is used, where IRX is the IR-to-UV luminosity ratio of galaxies, a robust indicator of UV dust attenuation $A_\mathrm{UV}$, and $\beta$ is the UV spectral slope.  Below are our conclusions.

\begin{enumerate}
	\item For the high-mass sample, the relation between IRX and $\beta$ is dependent on galaxy inclination. Edge-on galaxies ($b/a<0.5$, or inclination angle $i>62^\circ$) on average have greater IRX values by around 0.3 dex than face-on galaxies ($b/a>0.5$, or $i<62^\circ$) at the same UV spectral slope $\beta$. Edge-on galaxies also have lower UV luminosity than face-on galaxies at a given $\beta$. No dependence on galaxy inclination is found for the infrared luminosity $L$(IR). 
	
	\item To account for the inclination-dependence of the IRX--$\beta$ relation, we modify the \citet{Meurer1999} $A_\mathrm{UV}$ versus $\beta$ relation and obtain:
	$$
	A_\mathrm{UV,\, \mathrm{incl}}(\beta,\ b/a)=4.43+1.99\beta-1.73\,(b/a-0.67).
	$$
	This is consistent with studies of local massive star-forming galaxies, which show that the $A_\mathrm{UV}$ versus $\beta$ relation varies with galaxy inclination \citep{Battisti2017, Leslie2018}.
	The corresponding equation of SFR in the UV, following the calibration of \cite{Kennicutt2012}, is:
	$$
	\mathrm{SFR}_{\mathrm{UV,\,\beta,\, incl}}=1.72\times 10^{-10}L(1600\,\mbox{\AA})\cdot 10^{0.4\,A_{\mathrm{UV},\,\mathrm{incl}}}.
	$$
	\item For the high-mass sample, the inclination dependence of the IRX--$\beta$ relation indicates that edge-on galaxies have flatter (i.e., grayer) dust attenuation curves than face-on galaxies with the same line-of-sight dust attenuation $A_\mathrm{UV}$. Similarly, studies of local galaxies also find that UV dust attenuation curves become grayer at higher inclination in the same mass range \citep{Wild2011, Chevallard2013, Salim2018}.
		
	\item For the low-mass sample, galaxy UV luminosities at a given $\beta$ do not vary with inclination. They do at high mass. This indicates that low-mass galaxies at various inclinations likely have the same dust attenuation $A_\mathrm{UV}$ at a given $\beta$. Although their infrared luminosity $L$(IR) and the infrared excess IRX cannot be directly measured with the current dataset, as a corollary from the UV luminosity versus $\beta$ relation, their IRX--$\beta$ relation is not expected to have a dependence on inclination.
\end{enumerate}

To explain the inclination-dependence of the IRX--$\beta$ relation for high-mass galaxies, models of dust and star distributions from radiative transfer studies are adopted. A two-component dust model \citep{Charlot2000, Calzetti2001} is found to be the most plausible explanation. In this model, dust in the diffuse ISM forms a uniform mixture with stars (``mixture component''), and stars in \ion{H}{2} regions are additionally surrounded by local near-spherical dust shells (``shell component''). The relative contributions of the two components to the total dust attenuation vary with galaxy inclination.
At the same total UV dust optical depth, \emph{edge-on galaxies have a larger dust optical depth from the mixture component than face-on galaxies}, because the path-lengths through their disks are larger \citep{Wild2011,Battisti2017}. 
The observed UV light is dominated by young stars near the galaxy surface for the mixture component. These stars experience low dust reddening, outshining the other stars that are located deeper inside the galaxy disk, and therefore make the entire galaxy have blue UV spectrum (e.g., \citealt{Witt2000, Nordon2013, Seon2016,Popping2017}). This effect is stronger for edge-on galaxies and causes them to have a bluer (smaller) $\beta$ than face-on galaxies at the same IRX value.

Low-mass galaxies have more prolate shapes, as measured from the distribution of stars \citep{VanderWel2014, Tomassetti2016}. Furthermore, gas kinematics of these galaxies is highly disturbed (e.g., \citealt{Simons2017}), indicating that the distribution of gas and dust can be highly irregular. In other words, low-mass galaxies do not have disk-like shapes and kinematics. This explains why their $A_\mathrm{UV}-$$\beta$ relation does not depend on inclination, whereas it does for more massive galaxies. 



\acknowledgments
We thank the anonymous referee for the very comprehensive and constructive comments on the paper. WW is grateful for support by SAK through a grant from the STScI Director's Discretionary Research Fund (DDRF). CP is grateful for support from NASA grant 15-ADAP15-0186. PGP-G wishes to acknowledge support from Spanish Government MINECO Grant AYA2015-63650-P.  

This work is based on observations taken by the CANDELS Multi-Cycle Treasury Program with the NASA/ESA \emph{HST}, which is operated by the Association of Universities for Research in Astronomy, Inc., under NASA contract NAS5-26555. This work is also based on observations made with the \emph{Spitzer} Space Telescope operated by and funded through the Jet Propulsion Laboratory, California Institute of Technology under a contract with NASA. This research made use of Astropy, a community-developed core Python package for Astronomy \citep{astropy2018}.
\appendix

\section{Uncertainties resulting from measuring $\beta$ from broadband photometry and the potential influence of the 2175\,\AA\ bump}
\label{sec:appendeix}
We evaluate systematic uncertainties in the measurement of UV spectral slopes $\beta$. In short, the $\beta$ values measured in this work are 0.1-0.3 redder than the original spectroscopically-defined values by \citet{Calzetti1994}, which we take to be the ``gold standard.'' This systematic offset does not correlate with galaxy inclination, unless a rest-frame 2175\,\AA\ bump in the dust attenuation curve is present and its amplitude varies with galaxy inclination. The 2175\,\AA\ bump gives a systematic error of around 0.1 in the measurement of UV spectral slopes. Going back to Figure \ref{fig:irx_beta}, at fixed IRX, our photometrically-measured $\beta$ values of face-on galaxies are on average around 0.4 greater than those of edge-on galaxies.  The influence of the 2175\,\AA\ bump is not large enough to explain the total $\sim$0.4 difference. Details about the systematic offset follow.
\par 
\begin{figure}
	\figurenum{A1}
	\centering
	\includegraphics[width=5.in]{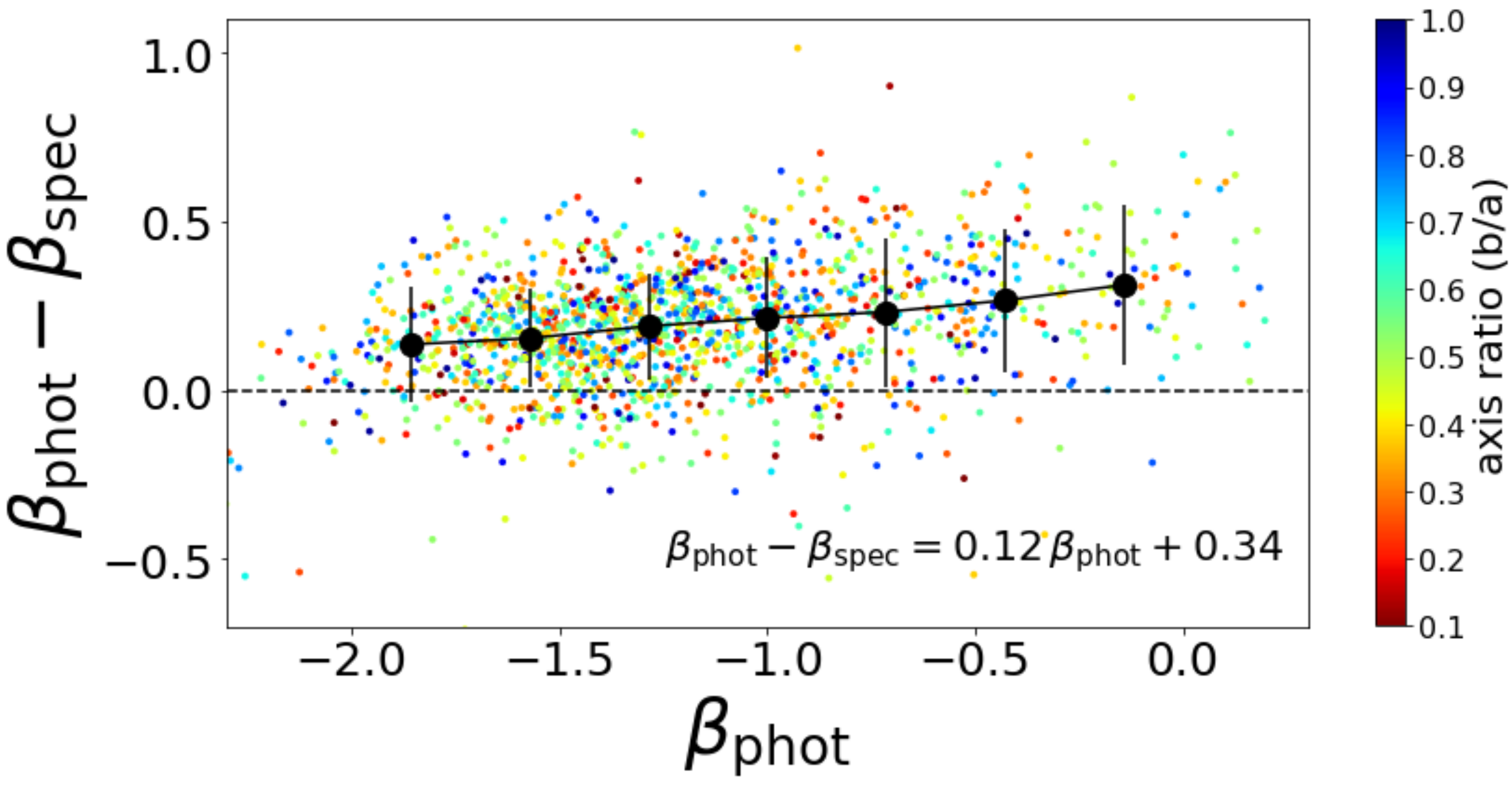}
	\caption{The UV spectral slopes $\beta_{\mathrm{phot}}$, as measured from broadband photometry in this work, are on average redder by 0.2-0.3 than the spectroscopically defined values, $\beta_{\mathrm{spec}}$. This is because the two measurements sample different parts of the UV spectrum. 
		In spite of a systematic difference between $\beta_{\mathrm{phot}}$ and $\beta_{\mathrm{spec}}$, the difference is not correlated with galaxy axis ratio (b/a, indicated by the color bar), and does not influence the \emph{relative} offset between face-on and edge-on galaxies on the IRX--$\beta$ diagram.  Galaxies in the high- and low-mass samples are shown in the plot. The quantity $\beta_{\mathrm{spec}}$ is derived from best-fit spectra in the ten narrow UV wavelength windows, as defined by \citet{Calzetti1994}. The SED-fitting follows \citet{Pacifici2016}, which assumes the \citet{Charlot2000} dust model without the 2175\,\AA\ bump feature. 
		\label{fig:beta_sedfittingtest}}
\end{figure}

\begin{figure*}
	\figurenum{A2}
	\centering
	\includegraphics[width=4.0in]{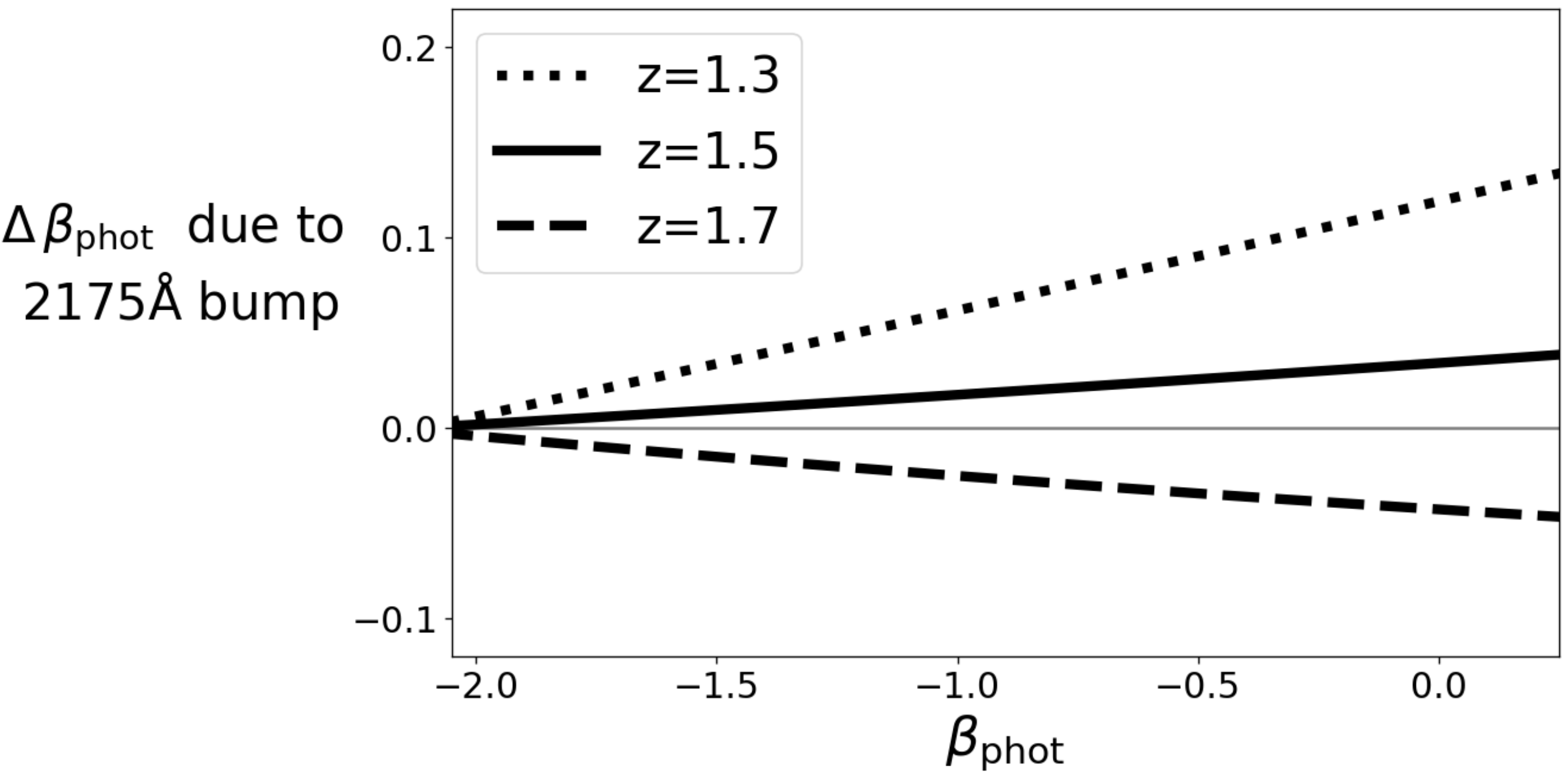}
	\caption{A potential 2175\,\AA\ bump feature in the dust attenuation curves influences our measured UV spectral slopes by no more than 0.1. The quantity $\Delta\,\beta_{\mathrm{phot}}$, which is the difference in measured UV slope values when the bump amplitude changes from zero to 50\% of that of the Milky-Way dust attenuation curve \citep{Cardelli1989}, is shown as a function of $\beta_{\mathrm{phot}}$. It correlates with galaxy inclination if face-on galaxies and edge-on galaxies have different 2175\,\AA\ bump strengths \citep{Kriek2013, Battisti2017}. The results shown here are based on the synthetic galaxy spectra from {\sc{galaxev}} model \citep{Bruzual2003} with a Milky-Way type dust attenuation. The UV slope $\beta_{\mathrm{phot}}$ is measured in the same wavelength ranges of the three broadbands used in the observations in this paper.
		\label{fig:deltabeta_beta}}
\end{figure*}

First, we examine whether our measurement of $\beta$ using broadband photometry deviates from other studies because different parts of the UV continuum are sampled. The UV slopes are calculated by fitting three broadband fluxes in this work ($\beta_{\mathrm{phot}}$, hereafter). In the seminal work of \cite{Calzetti1994}, they are measured from the ten narrower wavelength ranges of the UV spectrum ($\beta_{\mathrm{spec}}$, hereafter). The latter is taken as a standard among observational and theoretical studies (e.g., \citealt{Meurer1999,Narayanan2018}). To compare the two methods, we perform SED-fitting of the sample galaxies and measure $\beta_\mathrm{spec}$ from the best-fit model spectra. The SED-fitting is conducted on all the filters from U-band to IRAC 4.5\,\micron\ in the CANDELS catalog following \cite{Pacifici2016}, and assumes the \cite{Charlot2000} dust attenuation model which allows the dust attenuation curve shape to vary. This fitting does not include the 2175\,\AA\ bump feature in the dust attenuation curve. These $\beta_\mathrm{spec}$ values are compared to the UV slope values adopted in the main text, $\beta_\mathrm{phot}$. Figure \ref{fig:beta_sedfittingtest} shows that $\beta_\mathrm{phot}$ is on average 0.2-0.3 redder than $\beta_\mathrm{spec}$, with a random scatter of about 0.1. The difference between  $\beta_\mathrm{phot}$ and $\beta_\mathrm{spec}$ does not have any dependence on galaxy inclination, as indicated by the color-coding of Figure \ref{fig:beta_sedfittingtest}. Therefore, such difference does not influence the observed \emph{relative} offset between face-on and edge-on galaxies on the IRX--$\beta$ diagram (Figure \ref{fig:irx_beta}). 
\par 
Furthermore, we evaluate the influence of the 2175\,\AA\ bump on the UV slope measurement. When deriving UV spectral slopes in \S \ref{sec:measurement_beta}, we avoided the ACS/F606W band, which is close to the bump feature at $z\sim 1.5$. However, the presence of 2175 \AA\ bump may still have moderate influence on the other broadbands we use. The influence can be different for face-on and edge-on galaxies, if the strength of the 2175\,\AA\ bump systematically varies with galaxy inclination. A few recent studies at $0.5<z<3.0$ and in the local universe suggest that galaxies with higher inclination have stronger 2175\,\AA\ bump \citep{Kriek2013, Battisti2017}. A study by \citet{Tress2018} also found a weak inclination-dependence at $z\sim 2.0$. To test the influence of the 2175\,\AA\ bump, we generate the mock galaxy spectra with the stellar population synthesis code {\sc{galaxev}} \citep{Bruzual2003}, and measure  $\beta_{\mathrm{phot}}$ from the same broadband filters as used by our observations.  The spectrum modeling assumes a constant star-formation history, a mass-weighted age of 0.5 Gyr, a stellar metallicity of 0.2 $Z_{\sun}$, and the Milky-Way dust law \citep{Cardelli1989}. For the Milky-Way dust law, the 2175\,\AA\, bump can be either removed or reduced to 50\% of the original amplitude determined by \cite{Cardelli1989}, which resembles intermediate-redshift star-forming galaxies \citep{Kriek2013}. As shown in Figure \ref{fig:deltabeta_beta}, the change in the measured $\beta_\mathrm{phot}$ when switching from no bump to the 50\% bump is no more than 0.1. 
\par

\begin{figure*}
	\figurenum{B1}
	\centering
	\includegraphics[width=3.5in]{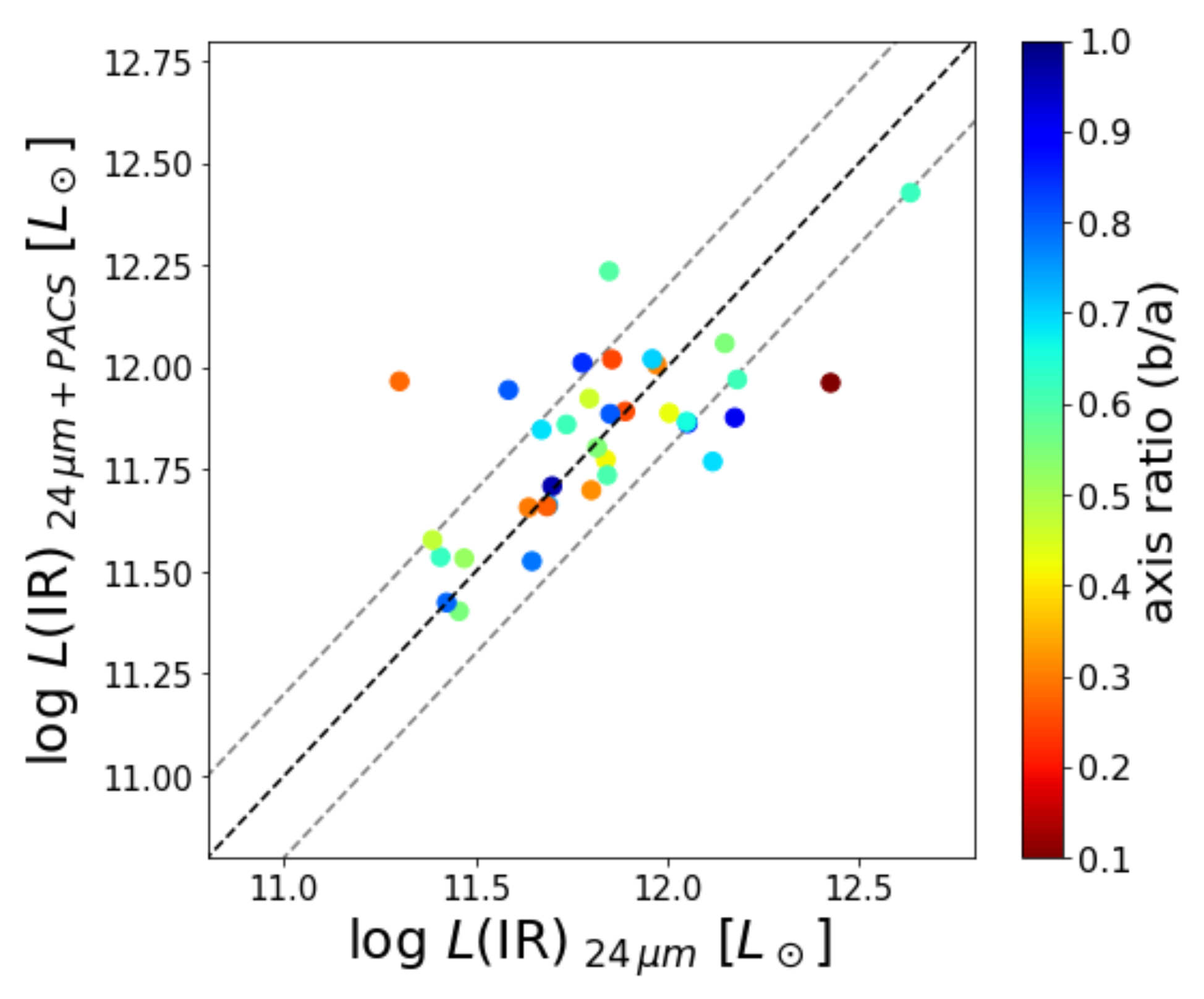}
	\caption{The total infrared luminosities inferred from \emph{Spitzer} MIPS/24\,\micron\ alone, $L$(IR)$_{\,24\,\micron}$, are consistent with those inferred from both 24\,\micron\ and the \emph{Herschel} far-IR bands (PACS/100\,\micron\ and PACS/160\,\micron), $L$(IR)$_{\,24\,\micron+\rm{PACS}}$. A total of 35 galaxies in our high-mass sample have detections in both the \emph{Spitzer} MIPS/24\,\micron\ band and the \emph{Herschel} far-IR bands, and are presented in the figure. The black dashed line marks the one-to-one relation. The difference between the two total infrared luminosities is typically no more than 0.2 dex, as marked by the gray dashed lines, and shows no dependence on galaxy axis ratio (color bar). $L$(IR)$_{\,24\,\micron+\rm{PACS}}$ is the median value as derived from four sets of SED templates \citep{Chary2001,Dale2001,Draine2007a,Rieke2009} by Barro et al.\ (in prep.).	\label{fig:herschelIR}}
\end{figure*}

\begin{figure*}
	\figurenum{C1}
	\centering
	\includegraphics[width=3.5in]{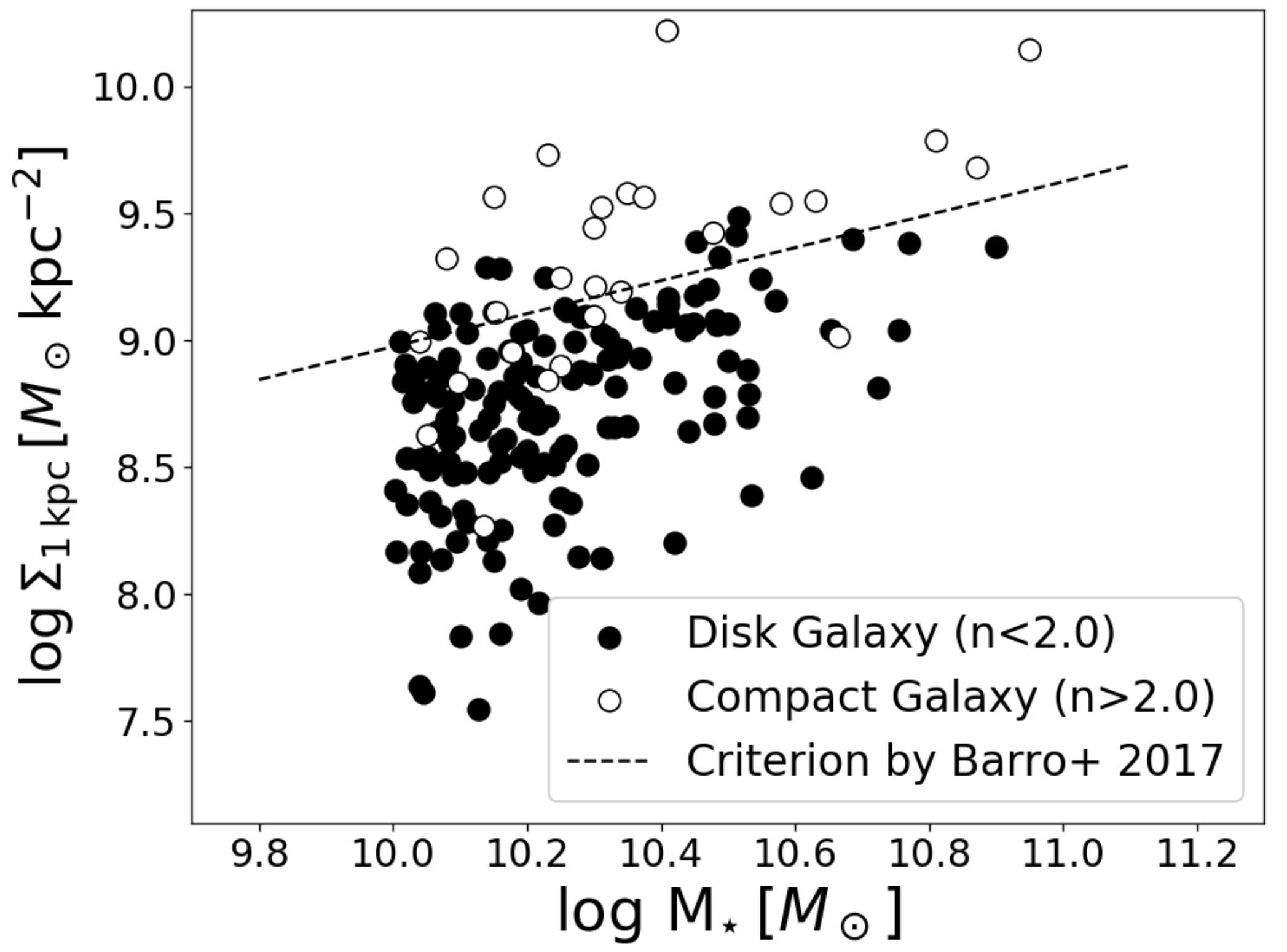}
	\caption{Two ways to identify compact galaxies are compared: selecting galaxies with H-band S\'ersic index $n>2$ (open symbols), as adopted in this work, and using the central stellar mass surface density ($\Sigma_\mathrm{1 kpc}$) versus stellar mass relation (dotted line, \citealt{Barro2017}). They are consistent with each other to first order. Galaxies with S\'ersic indexes greater than 2.0 generally lie above the selection cut by \cite{Barro2017}, whereas most of the galaxies with S\'ersic indexes smaller than 2.0 stay below. All the galaxies in the high-mass sample in addition to the massive star-forming galaxies without 24\,\micron\ fluxes are shown.
		\label{fig:sersic_hist}}
\end{figure*}
\section{Uncertainties In inferring $L$(IR) from $L$(24\,\micron)}
\label{sec:24umuncertainty} 
This appendix discusses systematic errors in inferring the total infrared luminosity $L$(IR) from the observed 24\micron\ luminosity alone. We argue that the rest-frame 9.7\,\micron\ silicate absorption feature is the only factor that gives an inclination-dependent error in the 24\,\micron-determined $L$(IR). This  inclination-dependent uncertainty is smaller than 0.1 dex and is not large enough to influence the main conclusions of this paper, i.e., that face-on galaxies and edge-on galaxies in the high-mass sample have \emph{relative} offsets on the IRX--$\beta$ diagram. A comparison between the 24\,\micron-determined $L$(IR) and the $L$(IR) derived with additional \emph{Herschel} far-IR photometry is also presented.
\par 
First, without considering infrared absorption features, galaxies should emit photons \textit{isotropically} in the mid- to far-infrared (e.g., \citealt{Jonsson2010,Leslie2018a}), and therefore the conversion from $L$(24\,\micron) to $L$(IR) should not contain any uncertainties that correlate with galaxy inclination. For our study, the only factor that may cause inclination-dependent uncertainties is the silicate absorption feature at rest-frame 9.7\,\micron. Radiative transfer results suggest that edge-on and face-on galaxies can have different amount of silicates along the line of sight, and this causes galaxies to have different mid-IR luminosities even if their far-IR luminosities are the same \citep{Jonsson2010}.  However, such an effect is only moderate according to recent observations. Observations of the local diffuse ISM and molecular clouds suggest that the optical depth at rest-frame 9.7\,\micron\  corresponds to approximately 1/20 of the visual attenuation $\,A_V$ (e.g., \citealt{VanBreemen2010}). The inclination-dependent uncertainty of $L$(IR) due to silicate absorption is at most 0.08 dex since the typical $A_V$ is around 1.5 mag for our high-mass galaxies. Such systematic uncertainty is not large enough to explain the difference in IRX between face-on and edge-on galaxies at the same $\beta$, which is around 0.3 dex.
\par 
Furthermore, we examine whether our method of inferring $L$(IR) from $L$(24\,\micron) is consistent with methods which use additional \emph{Herschel} photometry. The latter incorporate photometry in the far-IR, which is closer to the emission peak of galaxy infrared spectrum, and therefore may be able to constrain better the total infrared luminosity. A subsample of high-mass galaxies that have detections in \emph{Spitzer}/MIPS 24\micron, \emph{Herschel} PACS/100\,\micron, and PACS/160\,\micron\ are selected and shown in Figure \ref{fig:herschelIR}. New $L$(IR) values are obtained by fitting their mid-IR and far-IR luminosities to four sets of SED templates \citep{Chary2001,Dale2001,Draine2007a,Rieke2009}. For each galaxy, the median value of the four template fits is adopted as the total infrared luminosity, $L$(IR)$_{\,24\,\micron+\rm{PACS}}$. In Figure \ref{fig:herschelIR}, these values are compared with the 24\,\micron-determined total infrared luminosity, $L$(IR)$_{\,24\,\micron}$, and show a difference typically smaller than 0.2 dex. This agrees with recent studies of larger galaxy samples, which also find a similar scatter of around 0.2 dex \citep{Salmon2016}. In summary, for star-forming galaxies in the mass and redshift range studied in this paper, we expect that the $L$(IR) values derived from $L$(24\,\micron) fluxes are consistent to within 0.2 dex with those derived with additional \emph{Herschel} far-IR photometry.

\section{Identification of Compact Galaxies}
Though the majority of star-forming galaxies at $10.0<\,\log\,M_\star/M_\sun<11.0$ are disks, recent studies find a population of compact galaxies at this epoch (e.g., \citealt{Patel2012,VanderWel2014,Whitaker2015,Barro2017, Zhang2018}). In this work, we identify compact galaxies based on their S\'ersic indexes in the WFC3/F160W band.  Galaxies with S\'ersic indexes larger than 2.0 are defined as compact galaxies. Other studies use the stellar mass surface density within central 1 kpc, $\Sigma_{1\mathrm{kpc}}$, as an indicator of compact galaxies \citep{Barro2017}. We show in Figure \ref{fig:sersic_hist} that the two selection criteria are consistent and identify almost the same set of galaxies. Galaxies with high S\'ersic indexes also have high $\Sigma_{1\mathrm{kpc}}$ at a given stellar mass. 

As for the low-mass sample ($9.0<\,\log\,M_\star/M_\sun<10.0$), we do not identify compact galaxies in this paper (\S \ref{sec:measurement_incli}). A higher fraction (24.2\%) of the low-mass galaxies have S\'ersic indexes greater than 2.0, compared to 8.5\% for the high-mass sample. This is the evidence that low-mass galaxies do not have the same intrinsic shapes as high-mass galaxies (see also \citealt{VanderWel2014,Simons2017}). Nevertheless, the main result that low-mass galaxies do not have an inclination-dependent UV luminosity versus $\beta$ relation, as shown in Figure \ref{fig:umagvsb/a}, does not change even if objects with high S\'ersic indexes ($n>2.0$) are removed from the low-mass sample.
\label{sec:appendix_morp}
\bibliography{reference}
\bibliographystyle{aasjournal}
\end{document}